%% file: SIGIR2018.tex
\documentclass[sigconf]{acmart}

\usepackage{booktabs} 

\usepackage{algorithm}
\usepackage{tikz}
\usepackage{algpseudocode}
\usepackage{amssymb}
\usepackage{amsmath}
\usepackage{algorithmicx}
\usepackage{algpseudocode}
\usepackage{comment}
\usepackage{color}
\usepackage{url}
\usepackage{enumitem}
\usepackage{array}
\usepackage{multirow}
\usepackage[T1]{fontenc}

\newcommand{\mt}{\mathsf{T}}
\DeclareMathOperator*{\argmax}{arg\,max}

\def \bG {\mathbf{G}}

\begin{document}

\copyrightyear{2018} 
\acmYear{2018} 
\setcopyright{acmcopyright}
\acmConference[SIGIR '18]{The 41st International ACM SIGIR Conference on Research \& Development in Information Retrieval}{July 8--12, 2018}{Ann Arbor, MI, USA}
\acmBooktitle{SIGIR '18: The 41st International ACM SIGIR Conference on Research \& Development in Information Retrieval, July 8--12, 2018, Ann Arbor, MI, USA}
\acmPrice{15.00}
\acmDOI{10.1145/3209978.3210045}
\acmISBN{978-1-4503-5657-2/18/07}


\fancyhead{}  
\title{Efficient Exploration of Gradient Space \\for Online Learning to Rank}
\author{Huazheng Wang, Ramsey Langley, Sonwoo Kim, Eric McCord-Snook, Hongning Wang}
\affiliation{
  \institution{Department of Computer Science}
  \institution{University of Virginia} 
  \city{Charlottesville}
  \state{VA 22903}
  \country{USA}
 }
\email{{hw7ww,rml5tu,sak2m,esm7ky,hw5x}@virginia.edu}

\begin{abstract}

Online learning to rank (OL2R) optimizes the utility of returned search results based on implicit feedback gathered directly from users. To improve the estimates, OL2R algorithms examine one or more exploratory gradient directions and update the current ranker if a proposed one is preferred by users via an interleaved test. 

In this paper, we accelerate the online learning process by efficient exploration in the gradient space. Our algorithm, named as Null Space Gradient Descent, reduces the exploration space to only the \emph{null space} of recent poorly performing gradients. This prevents the algorithm from repeatedly exploring directions that have been discouraged by the most recent interactions with users. To improve sensitivity of the resulting interleaved test, we selectively construct candidate rankers to maximize the chance that they can be differentiated by candidate ranking documents in the current query; and we use historically difficult queries to identify the best ranker when tie occurs in comparing the rankers. Extensive experimental comparisons with the state-of-the-art OL2R algorithms on several public benchmarks confirmed the effectiveness of our proposal algorithm, especially in its fast learning convergence and promising ranking quality at an early stage.
\end{abstract}

\maketitle

\input{intro}

\input{related}
\input{method}
\input{exp}

\input{exp_analysis}

\section{Conclusions}
In this paper, we propose Null Space Gradient Descent (NSGD) to accelerate and improve online learning to rank. To avoid repeatedly exploring less promising directions, NSGD reduces its exploration space to the \emph{null space} of recently poorly performing directions. To identify the most effective exploratory rankers, NSGD uses a context-dependent preselection strategy to select candidate rankers that maximize the chance of being differentiated by an interleaved test for the current query. When two or more rankers tie, NSGD uses historically difficult queries to evaluate and identify the most effective ranker. We performed thorough experiments over multiple datasets and show that NSGD outperforms both the standard DBGD algorithm as well as several state-of-the-art OL2R algorithms. 


As our future work, it is important to study the theoretical properties of NSGD, including whether the directions proposed guarantee a low-bias estimation of the true gradients. As we observed in our empirical evaluations, the online trained models are generally worse than the offline trained ones, which benefit most from manual annotations. It is meaningful to combine these two types of learning schemes to maximize the utility of learnt models. Lastly, all OL2R algorithms consider consecutive interactions with users as independent; but this is not always true, particularly when users undergo complex search tasks. In this situation, balancing exploration and exploitation with respect to the search context becomes more important. We plan to explore this direction with our NSGD algorithm, as it already incorporates both long-term and short-term interaction history into gradient exploration. 

\section{Acknowledgments}
We thank the anonymous reviewers for their insightful comments. This work was supported in part by National Science Foundation Grant IIS-1553568 and IIS-1618948.

\bibliographystyle{ACM-Reference-Format}
\bibliography{sample-bibliography} 

\end{document}

%% file: intro.tex
\section{Introduction}

The goal of learning to rank is to optimize a parameterized ranking function such that documents that are more relevant to a user's query are ranked at higher positions \cite{liu2009learning}. A trained ranker combines hundreds of ranking features to recognize the relevance quality of a document to a query, and shows several advantages over the manually crafted ranking algorithms \cite{chapelle2011yahoo}. Traditionally, such a ranker is optimized in an offline manner over a manually curated search corpus. This learning scheme, however, becomes a main obstacle hampering the application of learning to rank algorithms for a few reasons: 1) it is expensive and time-consuming to obtain reliable annotations in large-scale retrieval systems; 2) editors' annotations do not necessarily align with actual users' preferences \cite{radlinski2008does}; and 3) it is difficult for an offline-trained model to reflect or capture ever-changing users' information needs in an online environment \cite{sanderson2010test}. 


To overcome these limitations, recent research has focused on learning the rankers on the fly, by directly exploiting implicit feedback from users via their interactions with the system \cite{joachims2002optimizing,chapelle2009dynamic,wang2014modeling}. Fundamentally, online learning to rank (OL2R) algorithms operate in an iterative manner: in every iteration, the algorithm examines one or more exploratory directions, and updates the ranker if a proposed one is preferred by the users via an interleaved test \cite{yue2009interactively,hofmann2013balancing,schuth2016multileave,zhao2016constructing}. The essence of this type of OL2R algorithms is to estimate the gradient of an unknown objective function with low bias, such that online gradient descent can be used for optimization with low regret \cite{flaxman2005online}. For example, one eventually finds a close to optimal ranker and seldom shows clearly bad results in the process. In the web search scenario, the objective function is usually considered to be the utility of search results, which can be depicted by ordinal comparisons in user feedback, such as clicks \cite{radlinski2008does}. However, to maintain an unbiased estimation of the gradient, \textit{uniform} sampling of random vectors in the entire parameter space is performed in these algorithms. As a result, the newly proposed exploratory rankers are \emph{independent} from not only the past interactions with users, but also the current query being served. This inevitably leads to slow convergence and large variance of ranking quality during the online learning process.

Several lines of works have been proposed to improve the algorithms' online learning efficiency. Hofmann et al. \cite{hofmann2013balancing} suggested to reduce the step size in gradient descent for better empirical performance. In their follow-up work \cite{hofmann2013reusing}, historical interactions were collected to supplement the interleaved test in the current query and pre-select the candidate rankers. Schuth et al. \cite{schuth2016multileave} proposed to explore multiple gradient directions in one multi-interleaved test \cite{schuth2014multileaved} so as to reduce the number of comparisons needed to evaluate the rankers. Zhao et al. \cite{zhao2016constructing} introduced the idea of using two uniformly sampled random vectors with opposite directions as the exploratory directions, with the hope that when they are not orthogonal to the optimal gradient, one of them should be a more effective direction than a simplely uniformly sampled direction. They also developed a contextual interleaving method, which considers historical explorations when interleaving the proposed rankers for comparison, to reduce the noise from multi-interleaving. 

Nevertheless, all aforementioned solutions still uniformly sample from the entire parameter space for gradient exploration. This results in independent and isolated rankers for comparison. Therefore, less promising directions might be repeatedly tested, as historical interactions are largely ignored when proposing the new rankers. More seriously, as the exploratory rankers are independently proposed for the current query, they might give the same ranking order of the candidate documents for interleaving (this happens when the difference in the feature weight vectors between two rankers are orthogonal to the feature vectors in those candidate documents). In this scenario, no click feedback can differentiate the ranking quality of those rankers in this query. When the interleaved test cannot recognize the best ranker from ordinal comparison in a query, tie will be arbitrarily broken \cite{yue2009interactively,schuth2016multileave}. This again leads to large variance and slow convergence of ranking quality in these types of algorithms.


We propose improving the learning convergence of OL2R algorithms by carefully exploring the gradient space. First, instead of uniformly sampling from the entire parameter space for gradient estimation, we maintain a collection of recently explored gradients that performed relatively poorly in their interleaved tests. 
We sample proposal directions from the \emph{null space} of these gradients to avoid repeatedly exploring poorly performing directions. Second, we use the candidate ranking documents associated with the current query to preselect the proposed rankers, with a focus on those that give \emph{different ranking orders} over the documents. This ensures that the resulting interleaved test will have a better chance of recognizing the difference between those rankers. Third, when an interleaved test fails to recognize the best ranker for a query, e.g., two or more rankers tie, we compare the tied rankers on the most recent worst performing queries (i.e., the \emph{difficult} queries) with the recorded clicks to differentiate their ranking quality. We name the resulting algorithm Null Space Gradient Descent, or NSGD for short, and extensively compare it with four state-of-the-art algorithms on five public benchmarks. The results confirm greatly improved learning efficiency in NSGD, with a remarkably fast and stable convergence rate at the early stage of the interactive learning process. This means systems equipped with NSGD can provide users with better search results much earlier, which is crucial for any interactive system.  


%% file: related.tex
\section{Related Work}

Online learning to rank has recently attracted increasing attention in the information retrieval community, as it eliminates the heavy dependency on manual relevance judgments for model training and directly estimates the utility of search results from user feedback on the fly. Various algorithms have been proposed, and they can be categorized into two main branches, depending on whether they estimate the utility of individual documents directly \cite{radlinski2008learning} or via a parameterized function over the ranking features \cite{yue2009interactively}.    

The first branch learns the best ranked list for each individual query by modeling user clicks using multi-armed bandit algorithms \cite{UCB1,Gambling}. 
Ranked bandits are studied in \cite{radlinski2008learning}, where a $k$-armed bandit model is placed on each ranking position of a fixed input query to estimate the utility of candidate documents being in that position. The system's learning is accelerated by assuming similar documents have similar utility for the same query \cite{slivkins2013ranked}. By assuming that skipped documents are less attractive than later clicked ones, Kveton et al. \cite{kveton2015cascading} develop a cascading bandit model to learn from both positive and negative feedback. To enable learning from multiple clicks in the same result ranking list, they adopt the dependent click model \cite{guo2009efficient} to infer user satisfaction after a sequence of clicks \cite{katariya2016dcm}, and later further extend to broader types of click models \cite{zoghi2017online}. However, such algorithms estimate the utility of ranked documents on a per-query basis, and no estimation is shared across queries. This causes them to suffer from slow convergence, making them less practical. 

Another branch of study leverages ranking features and look for the best ranker in the entire parametric space. Our work falls into this category. The most representative work in this line is dueling bandit gradient descent (DBGD) \cite{yue2009interactively, yue2012k}, where the algorithm proposes an exploratory direction in each iteration of interaction and uses an interleaved test to validate the exploration for model updating. 
As only one exploratory direction is compared in each iteration of DBGD, its learning efficiency is limited. Different solutions have been proposed to address this limitation. Schuth et al. \cite{schuth2016multileave} propose the Multileave Gradient Descent algorithm to explore multiple directions in each iteration. To evaluate multiple candidate rankers at once, multi-interleaving comparison \cite{schuth2014multileaved} is used. Zhao et al. \cite{zhao2016constructing} propose the Dual-Point Dueling Bandit Gradient Descent algorithm to sample two stochastic vectors with opposite directions as the candidate gradients. When they are not orthogonal to the optimal gradient, one of the two should be a more effective gradient than a single proposal. However, all of the aforementioned algorithms uniformly sample the exploratory directions from the entire parameter space, which is usually very high-dimensional. More importantly, the uniform sampling makes the proposed rankers independent from past interactions, and thus they cannot avoid repeatedly exploring less promising directions. 

Some works have recognized this deficiency and proposed different solutions. Hofmann et al. \cite{hofmann2013reusing} record historical interactions during the learning process to supplement the interleaved test when comparing the rankers. They also suggest using historical data to preselect the proposed rankers before interleaved test. However, only the most recent interactions are collected in these two solutions, so that they are not necessarily effective in recognizing the quality of different rankers. Oosterhuis et al. \cite{oosterhuis2017balancing} create the exploratory directions via a weighted combination over a set of preselected reference documents from an offline training corpus. The reference documents are either uniformly sampled or are the clustering centroids of the corpus. However, the reference documents are fixed beforehand; this limits the quality of learnt rankers, if the offline corpus has a different feature distribution than the incoming online documents. More importantly, none of these solutions consider the feature distribution in the candidate ranking documents of a particular query when proposing exploratory rankers. It is possible that the proposed rankers are not differentiable by any click pattern for a given query, e.g., they rank the documents in the same order. When the best rankers are tied, the winner is arbitrarily chosen. This further slows down the online learning process. In our solution, we preselect the rankers that tend to provide different ranking lists in the current query, so that the resulting interleaved test will have a better chance to tell the difference among those rankers. When a tie occurs, we use the most recent difficult queries to further evaluate the rankers, as those queries are expected to be more discriminative. 




%% file: method.tex
\section{Method}
\begin{figure*}[ht]
\centering
\setlength\tabcolsep{1pt}
\vspace{-1mm}
\begin{tabular}{c}
\hspace*{0cm}
\includegraphics[width=16.5cm]{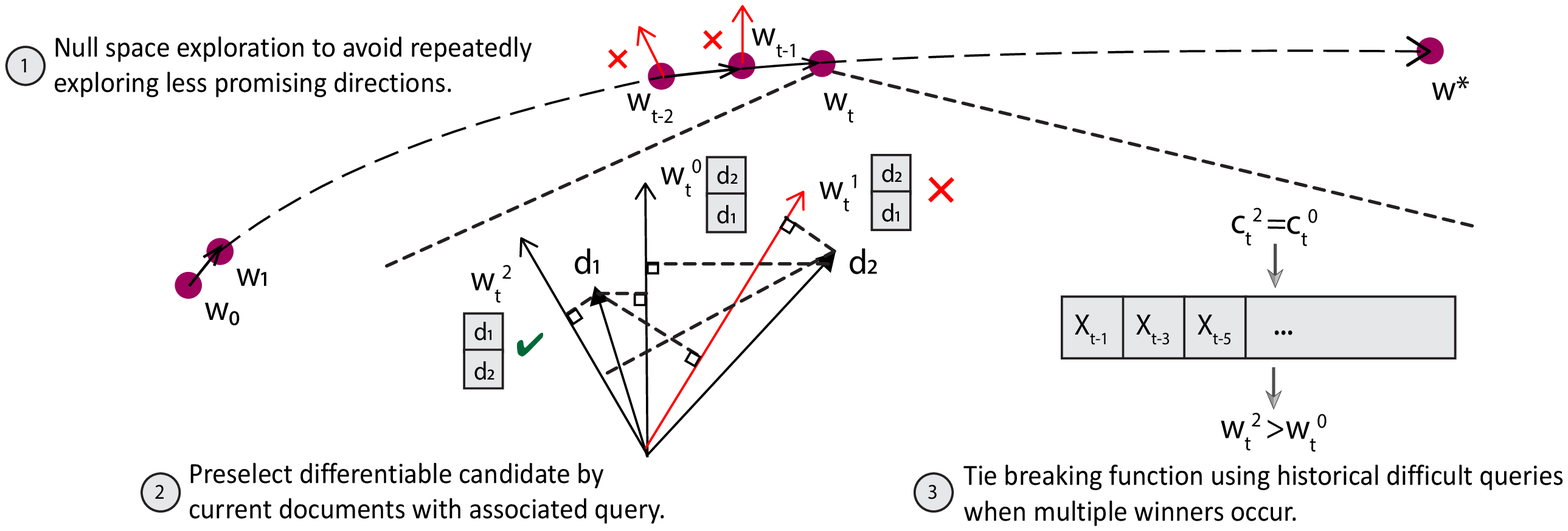} \\
\end{tabular}
\vspace{-2mm}
\caption{Illustration of model update procedure for the Null Space Gradient Descent algorithm.} \label{Fig:procedure}
\vspace{-2mm}
\end{figure*}

We improve the learning convergence of OL2R algorithms by carefully exploring the gradient space. In particular, we aim to avoid repeatedly exploring recent poorly performing directions and focus on the rankers that can be best differentiated by the candidate ranking documents associated with the current query.
We first give an overview of a basic online learning to rank algorithm, Dueling Bandit Gradient Descent \cite{yue2009interactively}, based on which we describe our proposed solution, Null Space Gradient Descent, in details. 

\subsection{Overview of Dueling Bandit Gradient Descent}
Dueling bandit gradient descent (DBGD) \cite{yue2009interactively} is an OL2R algorithm that learns from interleaved comparisons between one exploratory ranker and one current ranker. Each ranker is represented as a feature weight vector $w\in \mathbb{R}^d$, and it ranks documents by taking the inner product with their associated ranking features, i.e., a linear ranking model. As shown in Algorithm \ref{AlgDBGD}, at the beginning of iteration $t$, the algorithm receives a a query and its associated candidate ranking documents, represented as a set of query-document pairs $X_t = \{x_1, x_2, ..., x_s\}$. We denote $w_t^0$ as the weight vector of the current ranker. DBGD proposes an exploratory direction $u_t$ \textit{uniformly} from the unit sphere, and generates a candidate ranker $w_t^1 = w_t^0 + \delta u_t$, where $\delta$ is the step size of exploration. Two ranked lists generated by these two rankers, i.e., $l(X_t, w_t^0)$ and $l(X_t, w_t^1)$, are then combined via an interleaving method, such as Team Draft Interleaving \cite{radlinski2008does}. The resultant list is returned to the user for feedback. Based on the feedback and specific interleaving method, a better ranker is determined. If the exploratory ranker wins, the current ranker gets updated with $w_{t+1}^0 = w_t^0 + \alpha u_t$, where $\alpha$ is the learning rate; otherwise the current ranker stays intact. Such exploration and comparison lead to a low bias estimation of gradient in terms of expectation \cite{flaxman2005online}, i.e., $\nabla\hat f(w)=\mathbf{E}[f(w+\delta u)u]d/\delta$, in which $f(w)$ is the target utility function to be estimated. This estimation does not require the function $f(w)$ to be differentiable nor even to be explicitly defined; and thus it is the theoretical basis of this family of OL2R algorithms.

However, only one exploration direction $u_t$ is proposed for comparison in each iteration, which limits the learning rate of DBGD. To address this limitation, Schuth et al. \cite{schuth2014multileaved} proposed the Multileaving Gradient Descent algorithm that uniformly explores $m$ directions at the same time, i.e., ${w_t^0 + \delta u_t^i}$ where $i \in \{1..m\}$. Zhao et al. \cite{zhao2016constructing} proposed the Dual-Point Dueling Bandit Gradient Descent algorithm to explore two opposite directions each time, i.e., $w_t^0 + \delta u_t$ and $w_t^0 - \delta u_t$. Although exploring multiple candidates generally improves the learning rate, the expected improvement is still marginal, as the ranking features usually reside in a high dimensional space and uniform sampling is very inefficient. More importantly, the uniform sampling makes the proposed rankers independent from historical interactions and the current query context. Algorithms with history-independent exploration cannot avoid repeatedly exploring less promising directions that have been discouraged by the most recent user feedback. Additionally, context-independent exploration cannot avoid the issue of multiple rankers generating indifferentiable ranking results, such as by ranking the documents in the same order. Both of them further hamper the convergence rate of aforementioned OL2R algorithms. 

\begin{algorithm}[t]
\caption{Dueling Bandit Gradient Descent (DBGD) \cite{yue2009interactively}} \label{AlgDBGD}
\begin{algorithmic}[1]
\State \textbf{Inputs: } $\delta, \alpha$
\State Initiate $w_1^0 = \textit{sample\_unit\_vector}()$
\For{ $t=1$ to $T$}	
\State Receive query $X_t = \{x_1, x_2, ..., x_s\}$
\State $u_t = \textit{sample\_unit\_vector}() $
\State $w_t^1 = w_t^0 + \delta u_t$
\State Generate ranked lists $l(X_t, w_t^0)$ and $l(X_t, w_t^1)$
\State Set $L=$Interleave$\big(l(X_t, w_t^0), l(X_t, w_t^1)\big)$ and present $L$ to user
\State Receive click positions $C_t$ on $L$, and infer click credit $c_t^0$ and $c_t^1$ for $w_{t}^0$ and $w_{t}^1$ accordingly
\If { $c_t^1> c_t^0$}
\State {$w_{t+1}^0 = w_t^0 +\alpha u_t$}
\Else
\State $w_{t+1}^0 = w_t^0$
\EndIf
\EndFor
\end{algorithmic}
\end{algorithm}

\subsection{Null Space Gradient Descent} \label{sec:nullspace}

\begin{algorithm}[t]
\caption{Null Space Gradient Descent (NSGD)} \label{AlgNSGD}
\begin{algorithmic}[1]
\State \textbf{Inputs: } $\delta, \alpha, n, m, k_g, k_h, T_g, T_h$
\State Initiate $w_1^0 = \textit{sample\_unit\_vector}()$
\State Set $Q_g=queue(T_g)$ and $Q_h=queue(T_h)$ as fixed size queues
\For{ $t=1$ to $T$}	
\State Receive query $X_t = \{x_1, x_2, ..., x_s\}$
\State Generate ranked list $l(X_t, w_t^0)$
\State $\bar x_t = \sum_{i=1}^s x_i$
\State Construct $\bG = [g^{1}, ..., g^{k_g}]$ by directions selected from $Q_g$ with the worst recorded quality $q$
\State $\bG^\perp = NullSpace(\bG)$
\For {$i=1$ to $n$}
\State $g_t^i = \textit{sample\_unit\_vector}(\bG^\perp)$
\EndFor
\State Select top $m$ gradients that maximize $\left| \bar x_t^\mt {g_t^i} \right|$ from $\{g_t^i\}_{i=1}^n$
\For {$i=1$ to $m$}
\State $w_t^i =  w_t^0 + \delta g_t^i$
\State Generate ranked list $l(X_t, w_t^i)$
\EndFor
\State Set $L_t=\text{Multileave}\big(\{l(X_t, w_t^i)\}^m_{i=0}\big)$, and present $L_t$ to user
\State Receive click positions $C_t$ on $L_t$, and infer click credits ${\{c_t^i\}_{i=0}^m}$ for all rankers
\State Infer winner set $B_t$ from ${\{c_t^i\}_{i=0}^m}$
\If { $|B_t| > 1$ } 
\State Select $k_h$ worst performing queries $\big\{(X_i, L_i, C_i)\big\}_{i=1}^{k_h}$ from $Q_h$ by $\textit{Eval}(L_i, C_i) $.
\State $j  = \argmax_{o\in B_t} \sum_{i=1}^{k_h} \textit{Eval}(l(X_i, w_o), C_i) $
\Else
\State Set $j$ to the sole winner in $B_t$
\EndIf
\If {  $j = 0$ } 
\State $w_{t+1}^0 = w_t^0$
\Else
\State $w_{t+1}^0 = w_t^0 +\alpha g_t^j$
\EndIf
\For {$i=1$ to $m$}
\State $q_t^i = c_t^i - c_t^0$
\If {$q_t^i< 0$}
\State Append $(g_t^i, q_t^i)$ to $Q_g$
\EndIf
\EndFor
\State Append $(X_t, L_t, C_t)$ to $Q_h$
\EndFor
\end{algorithmic}
\end{algorithm} 

Our proposed Null Space Gradient Descent (NSGD) algorithm improves over DBGD-type OL2R algorithms by a suite of carefully designed exploration strategies in the gradient space. 

We illustrate the procedure of NSGD in Figure \ref{Fig:procedure}. First, to avoid uniformly testing exploratory directions in the entire parameter space, we maintain a collection of most recently explored gradients that performed poorly in their interleaved tests, and sample new proposal directions from the \emph{null space} of these gradients. As a result, we only search in a subspace that is orthogonal to those less promising directions. This can be intuitively understood from Figure \ref{Fig:procedure} part 1: since the interleaved tests in iteration $t-2$ and $t-1$ unveil the ineffectiveness of the directions marked in red, NSGD prevents the current ranker $w_t$ from exploring these less promising directions again by  exploring the null space of them at iteration $t$. Second, we prefer the proposed rankers that tend to generate \emph{the most distinct ranking orders} from the current ranker for the current query, so that the resulting interleaved test will have a better chance of recognizing the best ranker among those being compared. We show such an example in Figure \ref{Fig:procedure} part 2, where the current ranker $w_t^0$ and a randomly sampled candidate ranker $w_t^1$ rank the candidate documents in the same order. As a result, no interleaved test can differentiate their ranking quality in this query. NSGD avoids proposing $w_t^1$, and favors $w_t^2$ as it ranks the documents in a reverse order and would therefore give the interleaved test a better chance of recognizing the difference between $w_t^0$ and $w_t^2$. Third, if an interleaved test fails to recognize the best ranker in a query, e.g., a tie is encountered as shown in Figure \ref{Fig:procedure} part 3, we compare the tied rankers on the most recent worst performing queries (i.e., the \emph{difficult} queries) with the recorded clicks to differentiate the rankers. Eventually, NSGD aims to reach $w^*$ with a minimal number of interactions as shown in Figure \ref{Fig:procedure}, i.e., faster convergence. The detailed procedures in NSGD are shown in Algorithm \ref{AlgNSGD}, and we will discuss the key steps of it in the following.
\noindent\textbf{$\bullet$ Null Space Gradient Exploration.} NSGD maintains a fixed size queue $Q_g$ of recently explored directions and their corresponding quality, i.e., $Q_g=\big\{(g^i, q^i)\big\}^{T_g}_{i=1}$, which is constructed in line 32 to 37 in Algorithm \ref{AlgNSGD}. We denote the quality $q^i$ of an explored direction $g^i$ as the received click credit difference between the corresponding exploratory ranker and the default ranker by then (i.e., line 33). Intuitively, $q^i$ measures the improvement in ranking quality contributed by the update direction $g^i$; when $q^i$ is negative, it suggests the direction $g^i$ cannot improve the current ranker, and therefore should be discouraged in future. To realize this, after receiving a user query, NSGD first constructs $\bG = [g^{1}, ..., g^{k_g}]$ by selecting the top $k_g$ worst performing historical directions from $Q_g$ (i.e., line 8), and then solves for the null space of $\bG$ denoted as $\bG^{\perp} = NullSpace(\bG)$ (i.e., line 9). The new exploratory directions are sampled from $\bG^{\perp}$ (i.e., line 15). Because every vector in the space of $\bG^{\perp}$ is orthogonal to all $k_g$ selected historical directions, those ineffective directions (and any linear combination of them) will not be tested in this query.



Our null space exploration strategy is based on two mild assumptions: queries are independent and identically distributed (i.i.d.), and the gradient of the target (unknown) utility function satisfies Lipschitz continuity, i.e., $||\nabla\hat f(w_1) - \nabla\hat f(w_{2})||\le \gamma ||w_1 - w_{2}||$, where $\gamma$ is a Lipschitz constant for the target utility function $f(w)$. The assumption that queries are i.i.d. is studied and widely adopted in existing learning to rank research \cite{lan2008query,lan2009generalization}. This assumption allows NSGD to compare gradient performance across queries and select $k_g$ worst performing gradients from previous queries. Lipschitz continuity assumption suggests similar rankers would share similar gradient fields for the same query. This assumption is mild and consistent with most of existing learning to rank algorithms \cite{liu2009learning,burges2010ranknet}. However, this assumption requires us to construct the null space from all historically explored directions whose associated rankers have a similar weight vector $w$ as the current ranker's. This is clearly infeasible in an online learning setting, as we would have to store the entire updating history and examine it in every iteration. In NSGD, because the learning rate $\alpha$ is set to be small, rankers with close temporal proximity will have similar feature weight vectors, and therefore share a similar gradient field. Hence, NSGD only maintains the most recently tested directions in $Q_g$, which approximates the Lipschitz continuity. In our empirical evaluation, we also tested the exhaustive solution, but aside from the significantly increased storage and time complexity, little ranking performance improvement was observed. This supports our construction of the null space in NSGD.

Another benefit of sampling from null space is that the search takes place in a reduced problem space. DBGD-type algorithms have to sample in the whole $d$-dimensional space, while NSGD only samples from a subspace of it, whose rank is at most $d-k_g$, when the top $k_g$ worst performing historical gradients are orthogonal to each other. This advantage is especially appealing when the dimension of ranking features is high, which is usually the case in practical learning to rank applications \cite{liu2009learning,chapelle2011yahoo}.



There are two ways to sample from the null space $\bG^\perp$ in NSGD (i.e., line 11): uniformly selecting the basis vectors of the null space or sampling random unit vectors inside the null space. Randomly selecting the basis vectors can maximize the coverage of sampled directions in the null space, as the basis vectors are linearly independent from each other. It improves the exploration efficiency in an early stage. Zhao et al. tested a similar idea in \cite{zhao2016constructing}, but they performed it over the entire parameter space. However, in the later stage of model update, the true gradients are usually concentrating in a specific region; continuing to select those independent basis vectors becomes less effective. Exploring linear combinations of those basis vectors, i.e., uniformly sampling inside the space, emerges as a better choice then. But directly sampling from the null space at the beginning might be less effective, as it tends to introduce smaller variance in proposing different directions. 

To take advantage of these two sampling schemes, we propose a hybrid sampling method in the null space: comparing with the winning ranker $w_{t-k}^0$ created in iteration $t-k$, if $||w_t^0 - w_{t-k}^0|| < 1-\epsilon$, we switch to sample random unit vectors in $\bG^\perp$; otherwise uniformly select the basis vectors of $\bG^\perp$. The intuition behind this switching control is that when the consecutive rankers become similar, it indicates the gradients have converged to a local optimal region, and a refined search is needed to identify the true gradient. Otherwise, the gradient direction has not been identified, and larger diversity is needed to accelerate the exploration of null space. Oosterhuis and Rijke \cite{oosterhuis2017balancing} also proposed a similar idea to detect model convergence and convert to more complex models when a simpler model has converged. But their conversion might not always be feasible, e.g., when no linear mapping between models exists; we only switch the sampling schemes for exploratory directions, which has no additional assumption about the model space. 


\noindent\textbf{$\bullet$ Context-Dependent Ranker Preselection.} NSGD selectively constructs the candidate rankers to maximize the chance that they can be \emph{differentiated} from the current best ranker $w_t$ 
in the interleaved tests. A straightforward solution is to select the rankers which give totally distinct ranking orders to that from $w^0_t$. But this clearly emphasizes too much the exploration of new directions, but ignores the exploitation of current best ranker. Especially in the later stage of model update when the current ranker can already provide satisfactory ranking results, a very distinct ranking indicates a higher risk of providing worse result quality.

To balance the needs for exploration and exploitation, we propose a Context-Dependent Preselection (CDP) criterion as shown in line 13 of Algorithm \ref{AlgNSGD}: after randomly sampling $n$ vectors from $\bG^{\perp}$, we select the top $m$ of them that maximize the inner product with the aggregate document feature vector $\bar x$ for query $X_t$. This can be understood as a necessary condition for having a proposed ranker that generates a different ranked list in $X_t$ than that from $w^0_t$. More specifically, as we are learning a linear ranker, the ranking score of each document is computed by the inner product between document feature vector $x_i$ and the feature weight vector $w^0_t$; and the ranking scores lead to the ranked list $l(X_t, w^0_t)$. To generate a different ranked list, there has to be at least one document that has different ranking scores under these two rankers, i.e., $\exists j, |x^\mt_j (w_t^i - w^0_t)| > 0$. This can be  simplified as $\exists j, |x^\mt_j g_t^i| > 0$; and by the triangle inequality (i.e., $|a| + |b| \ge |a+b| $), we require a differentiable ranker to satisfy $|\sum_j x_j^\mt g_t^i|>0$. To choose the candidate rankers that can best satisfy this condition, we select the top $m$ proposal directions that maximize this inner product. 


\noindent\textbf{$\bullet$ History-Dependent Tie Breaking.} NSGD is flexible in selecting the number of rankers for comparison: the hyper-parameter $m$ in line 13 is an input to the algorithm. If multiple rankers are selected for comparison, multi-interleaving \cite{schuth2014multileaved} can be performed to compare the quality of the proposed rankers, i.e., infer the click credit $c_t^i$ for each ranker $w_t^i$ and determine the winning ranker (i.e., line 19 and 20). However, because of position bias in user clicks \cite{joachims2017accurately}, very few  result documents will be clicked each time. The sparsity in result clicks directly reduces the resolution of interleaved test in recognizing the winning ranker, e.g., multiple rankers might share the same aggregate click credit. The situation becomes even worse when multiple rankers are compared. Existing solutions break the tie arbitrarily \cite{yue2009interactively,zhao2016constructing} or heuristically take the mean vector of rankers in the winner set \cite{schuth2016multileave}. No solutions consider the ranking problem at hand, and they are not effective in general.

We propose the idea of leveraging historical queries, especially the most difficult ones, to choose the winner whenever a tie happens. First, in line 38, the 3-tuple comprised of the historical query, its displayed ranking list, and its corresponding click positions are stored in a fixed size queue. In future iterations, they are selected in line 22 to identify the best ranker, whenever a tie happens. Because only click feedback is available in online learning, we use click position $C_t$ in the evaluation function $Eval(L_t,C_t)$, such as MAP or NDCG by treating clicked documents as relevant, to measure the ranking quality of $L_t$ in query $X_t$ (i.e., line 22 and 23). More importantly, because the ranker is improving on the fly, a poorly served query might be caused by a badly performing ranker, rather than its intrinsic difficulty. Therefore, in NSDG we only collect recent click results to select the most discriminative queries. 

%% file: exp.tex
\begin{figure*}[ht]
\centering
\setlength\tabcolsep{4pt}
\vspace{-1mm}
\begin{tabular}{ccc}
\hspace*{0cm}
\includegraphics[width=5.2cm]{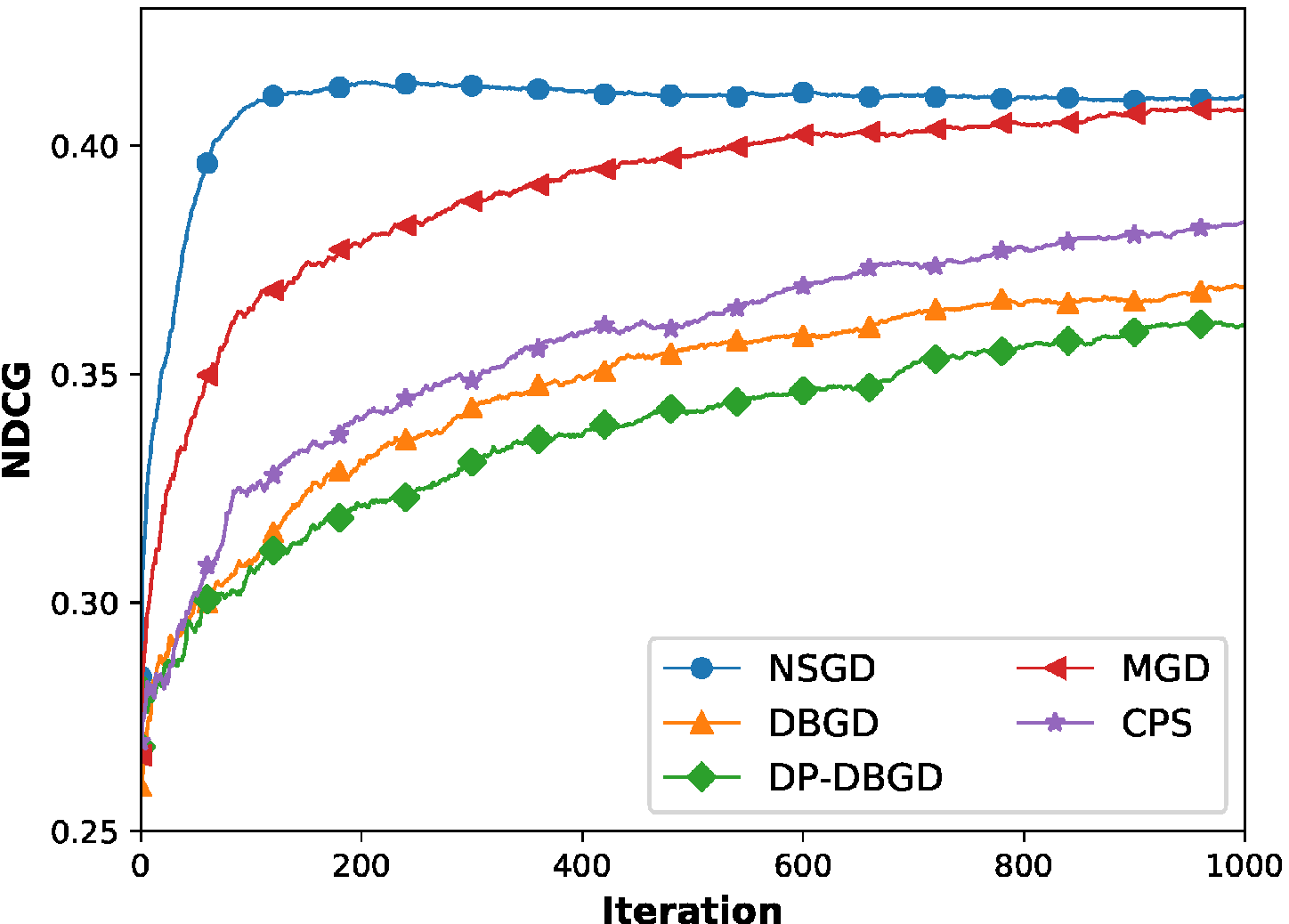} & 
\includegraphics[width=5.3cm]{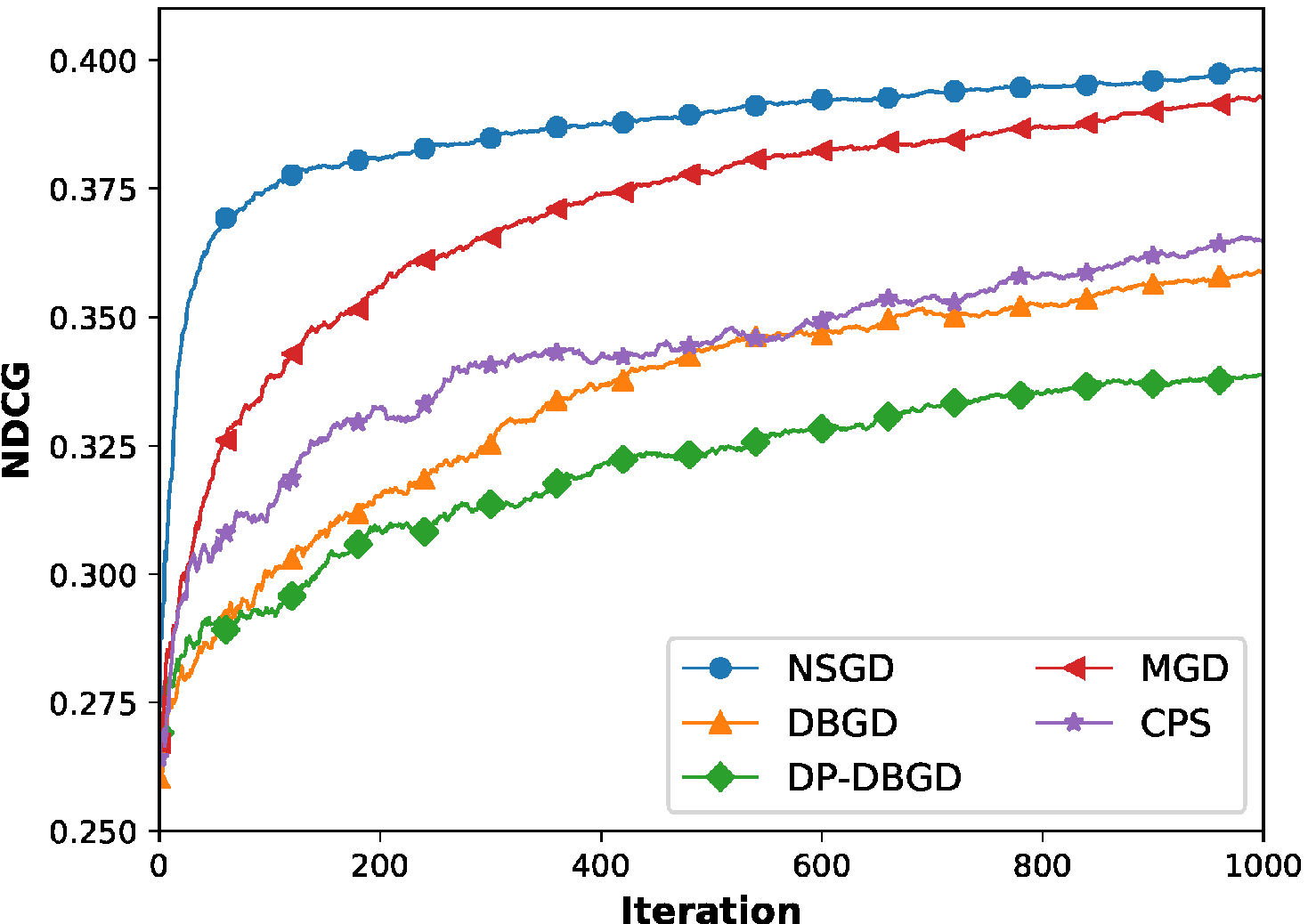} & 
\includegraphics[width=5.3cm]{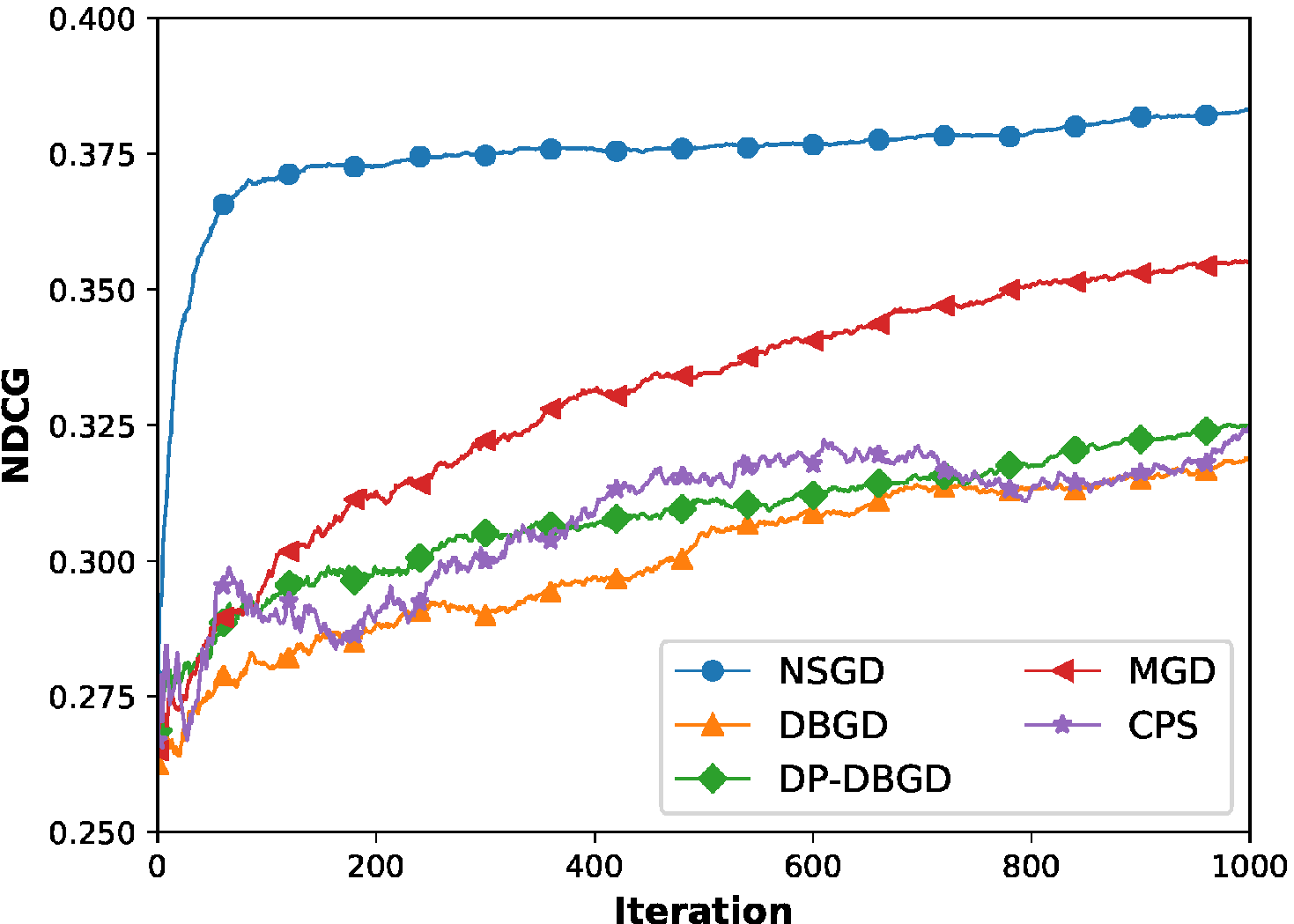} \\
(a) Perfect & (b) Navigational & (c) Informational \\
\end{tabular}
\vspace{-3mm}
\caption{Offline NDCG@10 on MQ2007 dataset under three click models.} \label{Fig:offline}
\vspace{-2mm}
\end{figure*}

\begin{figure*}[ht]
\centering
\setlength\tabcolsep{4pt}
\begin{tabular}{ccc}
\hspace*{0cm}
\includegraphics[width=5.3cm]{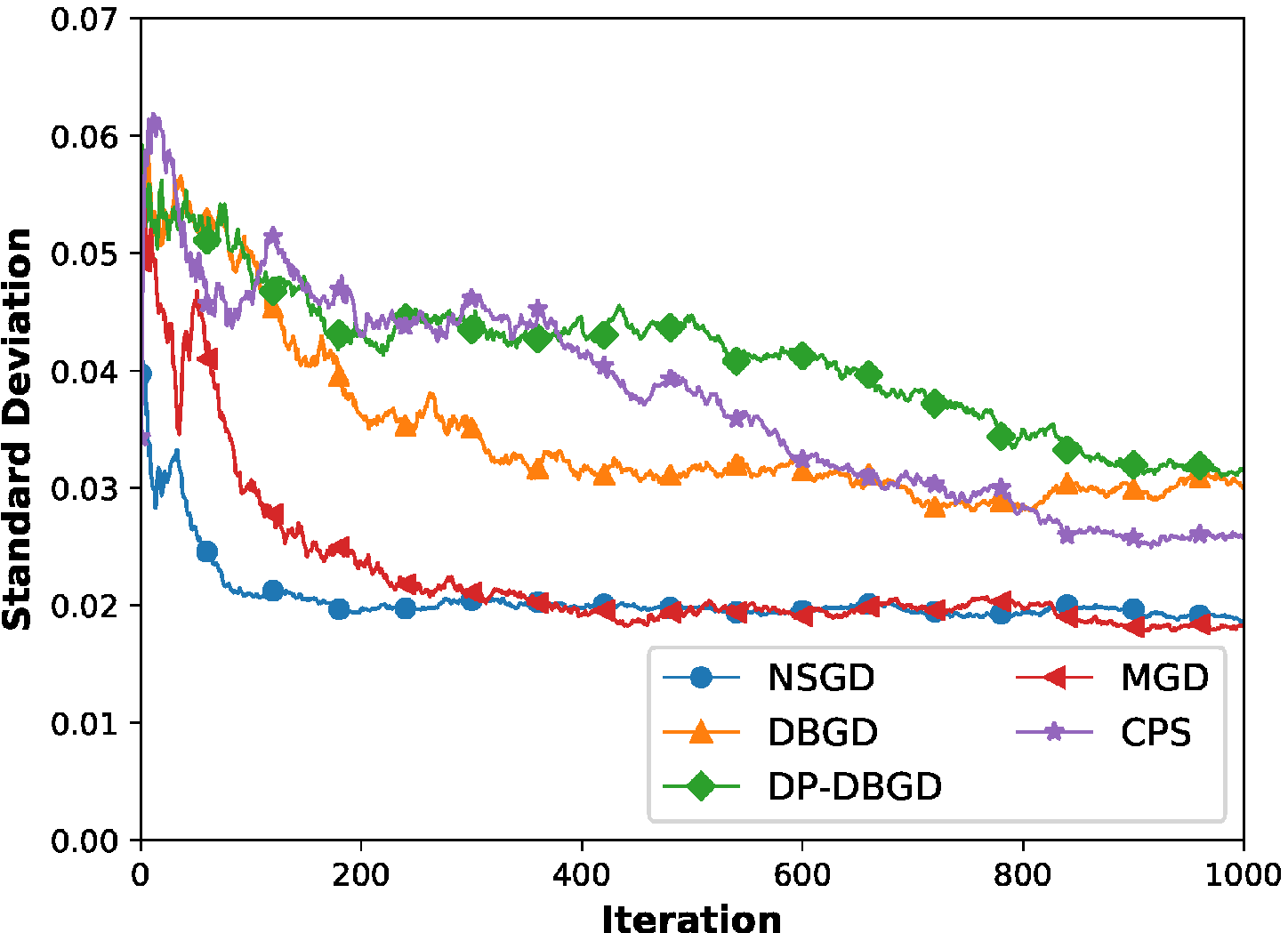} &
\includegraphics[width=5.3cm]{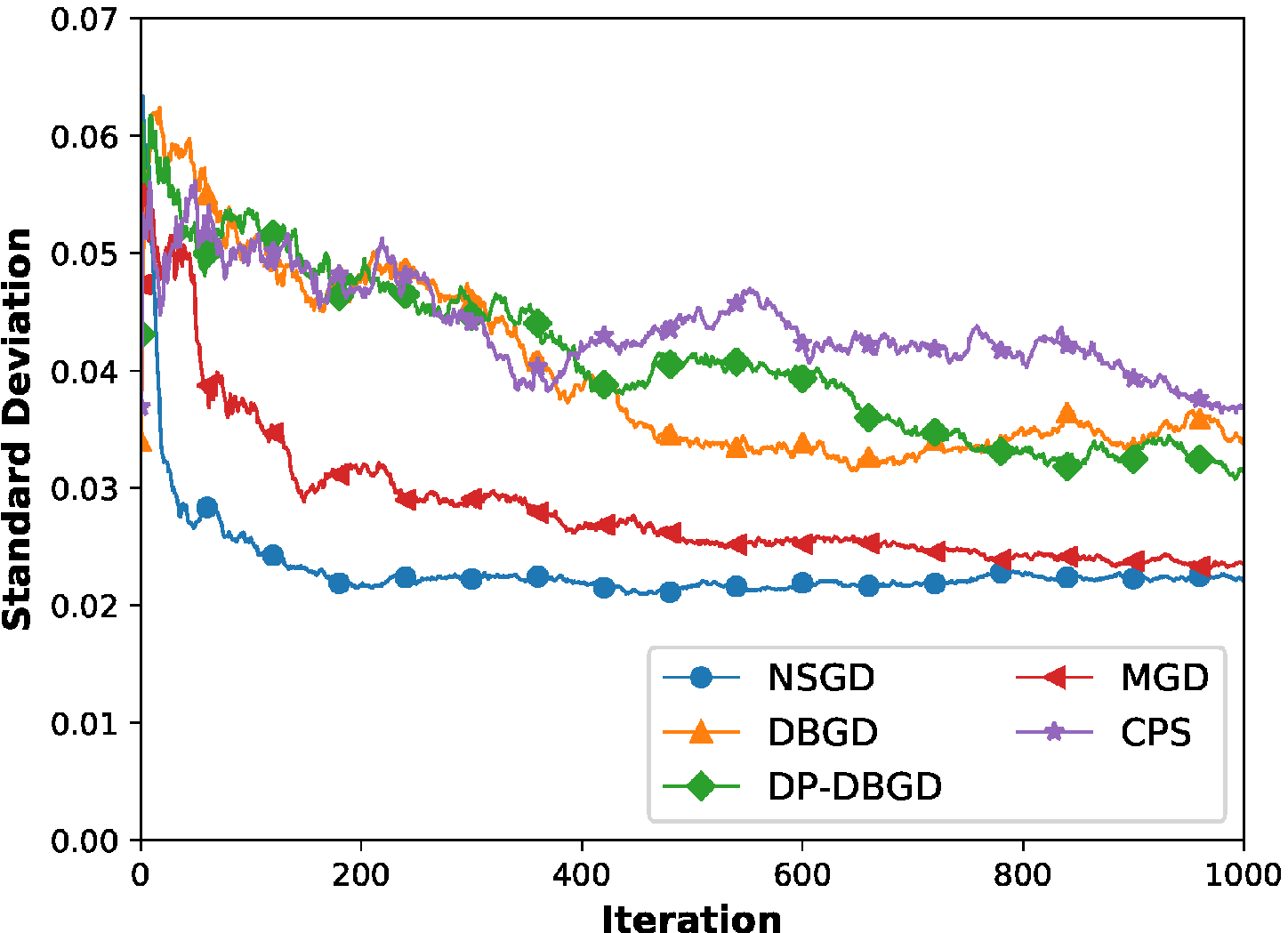} &
\includegraphics[width=5.3cm]{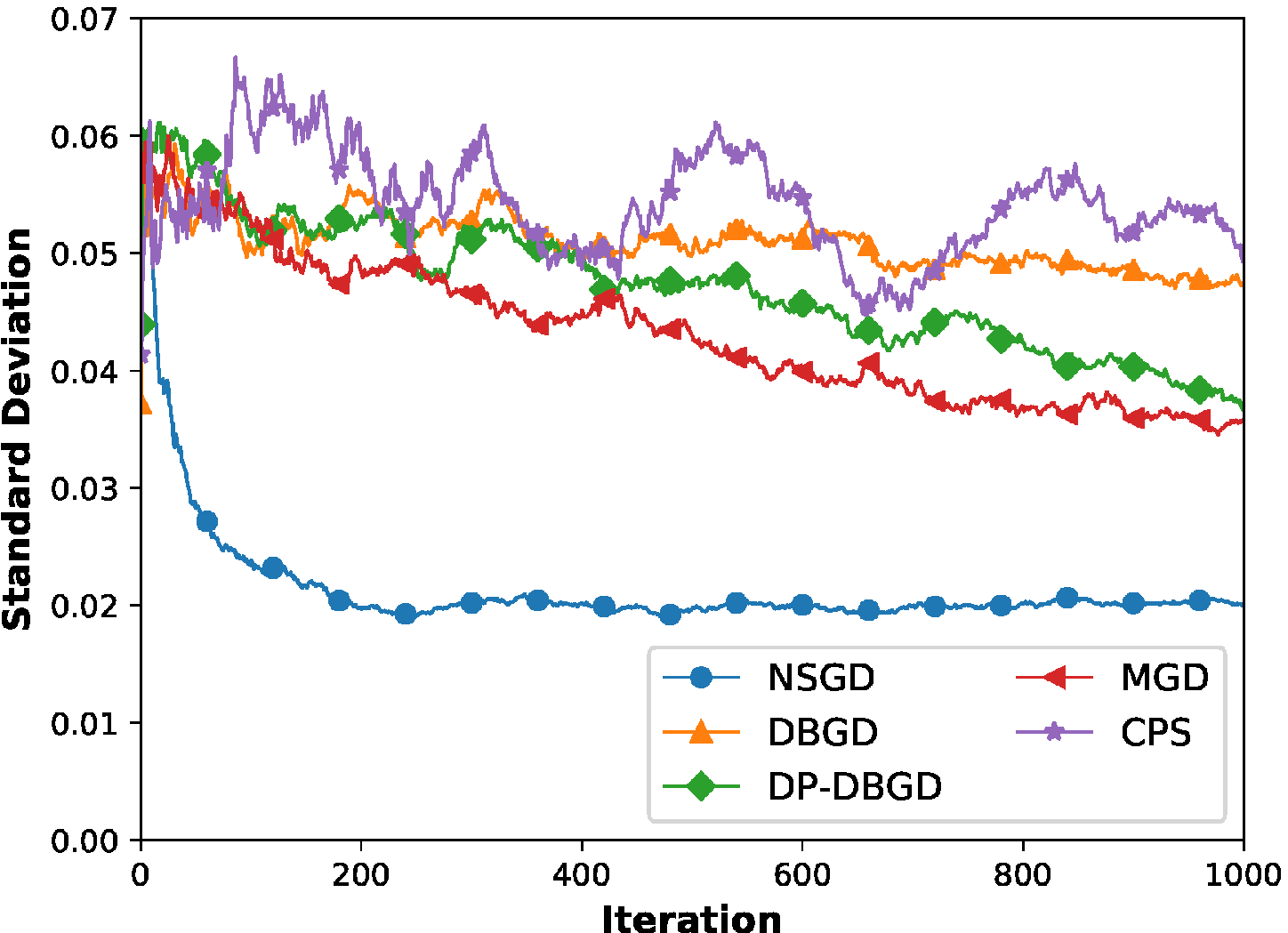} \\
(a) Perfect & (b) Navigational & (c) Informational \\
\end{tabular}
\vspace{-3mm}
\caption{Standard deviation of offline NDCG@10 on MQ2007 dataset under three click models.} \label{Fig:std}
\vspace{-2mm}
\end{figure*}

\begin{figure*}[ht]
\centering
\setlength\tabcolsep{4pt}
\begin{tabular}{ccc}
\hspace*{0cm}
\includegraphics[width=5.25cm]{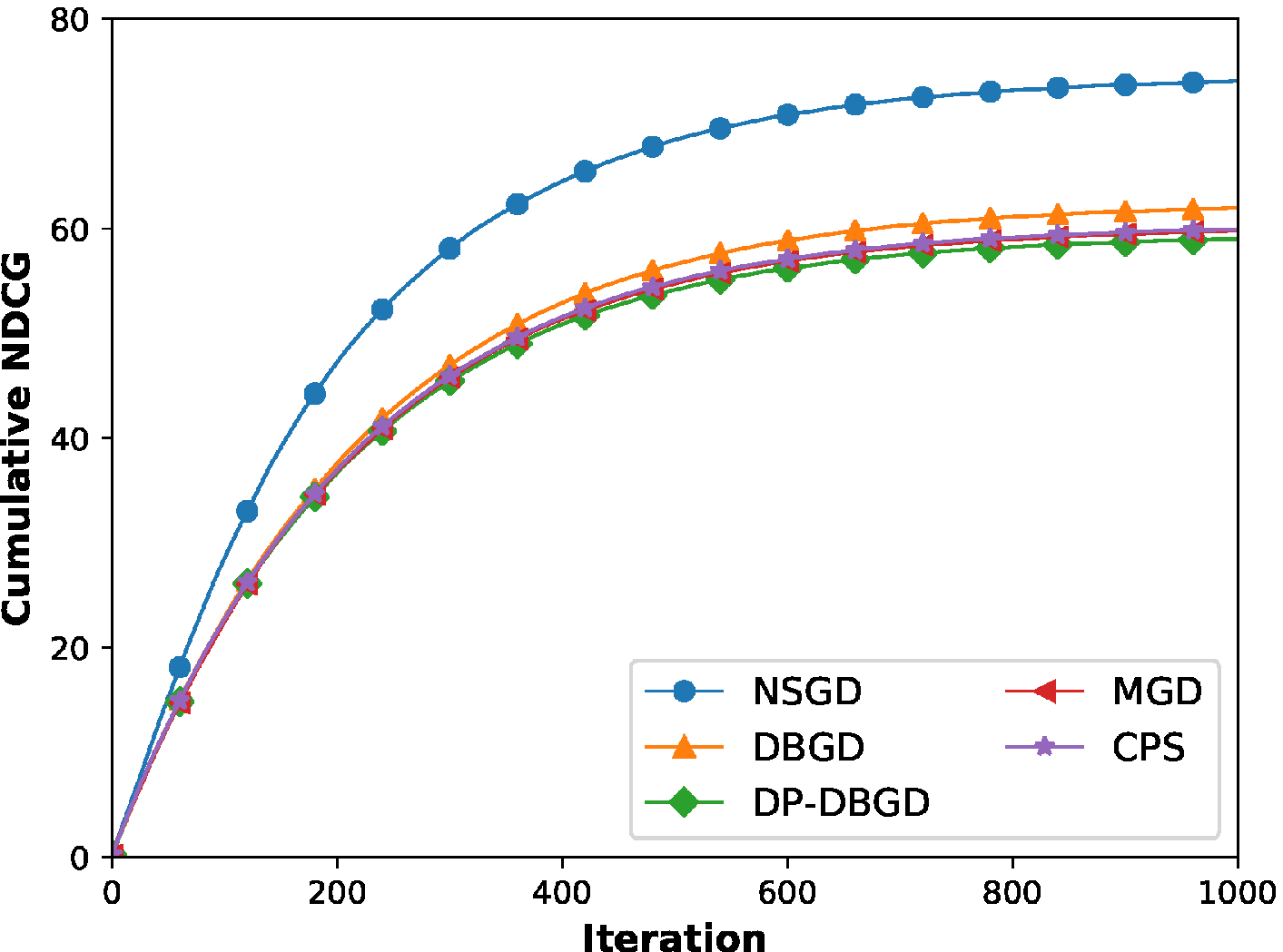} &
\includegraphics[width=5.25cm]{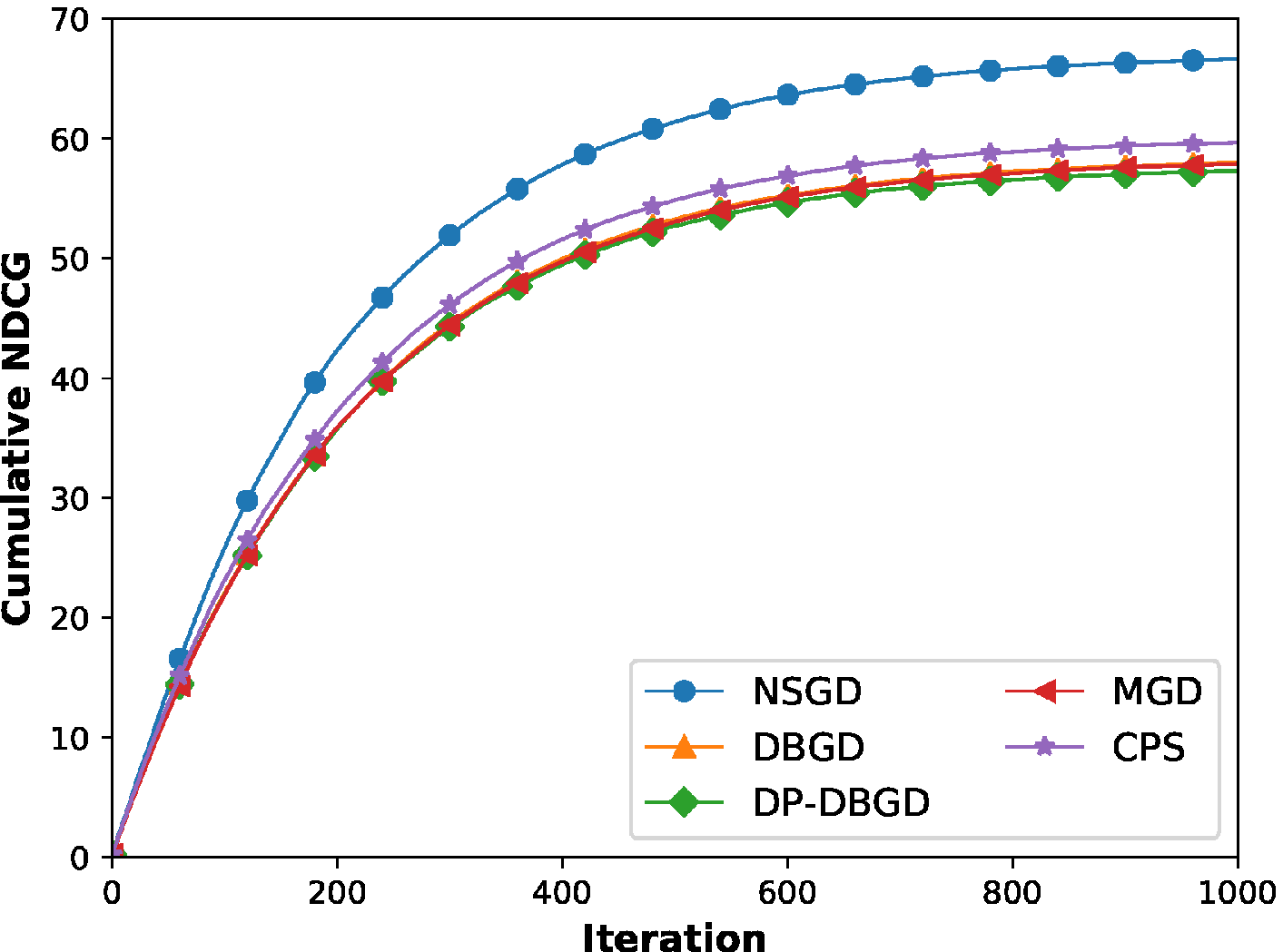} &
\includegraphics[width=5.25cm]{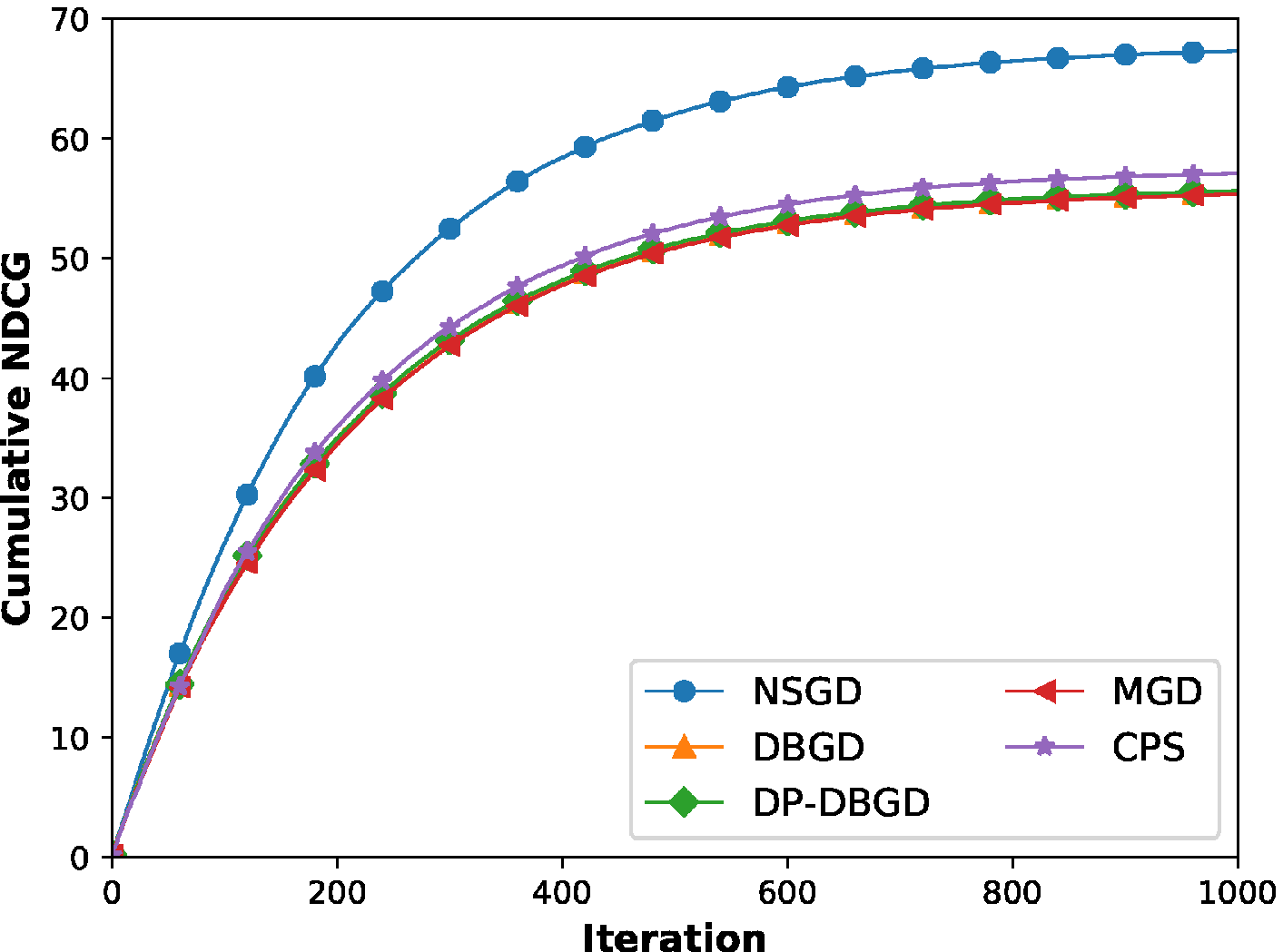} \\
(a) Perfect & (b) Navigational & (c) Informational \\
\end{tabular}
\vspace{-3mm}
\caption{Discounted cumulative NDCG@10 on MQ2007 dataset under three click models.} \label{Fig:cumulative}
\vspace{-4mm}
\end{figure*}

\section{Experiments}\label{sec:exp}
In this section, we perform extensive empirical comparisons between our proposed Null Space Gradient Descent (NSGD) algorithm with four state-of-the-art OL2R algorithms on five public learning to rank benchmarks. Both quantitative and qualitative evaluations are performed to examine our proposed gradient space exploration strategies, especially their advantages over the existing solutions in improving online learning efficiency. 

\subsection{Experiment Setup} \label{exp:setup}
\noindent\textbf{$\bullet$ Datasets.}
We used five benchmark datasets which are part of the LETOR 3.0 and LETOR 4.0 collections \cite{liu2007letor}: MQ2007, MQ2008, TD2003, NP2003 and HP2003. Among them, NP2003 and HP2003 implement navigational tasks, such as homepage finding and named-page finding; TD2003 implements topic distillation, which is an informational task; MQ2007 and MQ2008 mix both types of tasks. 
Documents in TD2003, NP2003 and HP2003 datasets are collected from the .GOV collection, which is crawled from the .gov domain; while the MQ2007 and MQ2008 datasets are collected from 2007 and 2008 Million Query track at TREC \cite{voorhees2005trec}. In these datasets, each query-document pair is encoded as a vector of ranking features, including PageRank, TF.IDF, BM25, and language model on different parts of a document. The number of features is 46 in MQ2007 and MQ2008, and 64 for the other three datasets. In the MQ2007 and MQ2008 datasets, every document is marked with a relevance label between 0 and 2, while the other datasets only have binary labels. The MQ2007 and MQ2008 datasets contain 1,700 and 1,800 queries respectively, but with fewer assessments per query; while each of the other three datasets only contain fewer than 150 queries but with 1,000 assessments per query. All of the datasets are split into 5 folds for cross validation. We take the training set for online experiments gathering cumulative performance, and use testing set for offline evaluation.

\noindent\textbf{$\bullet$ Simulating User Clicks.}
To make our reported results comparable to existing literature, we follow the standard offline evaluation scheme proposed in Lerot \cite{schuth2013lerot}, which simulates user interactions with an OL2R algorithm. We make use of the Cascade Click Model \cite{guo2009efficient} to simulate user click behavior. The click model simulates user interaction with the system by assuming that as a user scans through the list he/she makes a decision about whether or not to click on a returned document. The probability of a user clicks on a document is conditioned on the relevance label. Likewise, after clicking, the user makes a decision about continuing to look through the documents or to stop. The probability of this decision is also conditioned on the current document's relevance label. Adjusting these probabilities allows us to simulate different types of users. 

We use three click model configurations as shown in Table \ref{tab:clickmodel}, including: 1) \textit{perfect} user who clicks on all relevant documents and does not stop browsing, which contributes the least noise; 2) \textit{navigational} user who would stop early once they've found a relevant document; and 3) \textit{informational} user who sometimes clicks on irrelevant documents in their search for information, which contributes the most noise. The length of resulting list evaluated by the click models is set to 10 as a standard setting in \cite{schuth2016multileave,zhao2016constructing}.

\noindent\textbf{$\bullet$ Evaluation Metrics.}
To evaluate an OL2R algorithm, cumulative Normalized Discounted Cumulative Gain (NDCG) and offline NDCG are commonly used to assess the learning rate and ranking quality of the algorithm \cite{schuth2013lerot}. 
Cumulative NDCG is calculated with a discount factor of $\gamma$ set to 0.995 for each iteration. To assess model estimation convergence, in each iteration we measure cosine similarity between the weight vector updated by an OL2R algorithm and a reference weight vector, which is estimated by an offline learning to rank algorithm trained on the manual relevance judgments. In our experiment, we used LambdaRank \cite{burges2010ranknet} with no hidden layer to obtain such a reference ranker in each dataset, because of its superior empirical performance. For all experiments, we fix the total number of iterations $T$ to 1,000 and randomly sample query $X_t$ from the dataset with replacement accordingly.

\begin{table}[t]
    \centering
    \caption{Configurations of simulation click models.}\label{tab:clickmodel}
    \vspace{-3mm}
    \begin{tabular}{ccccccc}
    \hline
         & \multicolumn{3}{c}{Click Probability} & \multicolumn{3}{c}{Stop Probability} \\
        Relevance grade & 0 & 1 & 2 & 0 & 1 & 2\\
         \hline
        Perfect & 0.0 & 0.5 & 1.0 & 0.0 & 0.0 & 0.0\\
        Navigational & 0.05 & 0.5 & 0.95 & 0.2 & 0.5 & 0.9\\
        Informational & 0.4 & 0.7 & 0.9 & 0.1 & 0.3 & 0.5\\
        \hline
    \end{tabular}
    \vspace{-3mm}
\end{table}

\noindent\textbf{$\bullet$ Evaluation Questions.}
We intend to answer the following questions through empirical evaluations, to better understand the advantages of our proposed algorithm.
\begin{itemize}
\item[Q1:] How does our proposed NSGD algorithm perform in comparison to various baseline OL2R methods?
\item[Q2:] Do candidate directions generated by NSGD explore the gradient space more efficiently than uniform sampling from the entire parameter space?
\item[Q3:] How do the different components in NSGD contribute to its final performance?
\item[Q4:] How do different settings of hyper-parameters alter the performance of NSGD?
\end{itemize}

\noindent\textbf{$\bullet$ Baselines.}
We chose the following four state-of-the-art OL2R algorithms as our baselines for comparison:
\begin{itemize}
    \item[-] \textbf{DBGD}~\cite{yue2009interactively}: A single direction uniformly sampled from the entire parameter space is explored. Team Draft is used to interleave the results of the two rankers for comparison.
    \item[-] \textbf{CPS}~\cite{hofmann2013balancing}: It proposes a candidate preselection strategy that uses historical data to preselect the proposed rankers before the interleaved test in DBGD.
    \item[-] \textbf{DP-DBGD}~\cite{zhao2016constructing}: Two opposite uniformly sampled directions are explored in DBGD. Both Contextual Interleave, which favors the winning direction from the previous iteration, and Team Draft are used in it in our experiment.
    \item[-] \textbf{MGD}~\cite{schuth2016multileave}: Multiple uniformly sampled directions are explored in single iteration. Multileave is used to interleave the results. If there is a tie, the model updates towards the mean of all winners.
\end{itemize}


%% file: exp_analysis.tex


\begin{table*}[ht]
    \centering
    \vspace{-1mm}
    \caption{Online score (discounted cumulative NDCG@10) and standard deviation of each algorithm after 1000 queries under each of the three click models. Statistically significant improvements over MGD baseline are indicated by $\blacktriangle$ (p<0.05).}
    \vspace{-2mm}
    \begin{tabular}{cccccccccc}
        \hline
    Click Model   &Dataset& DBGD & CPS & DP-DBGD & MGD  & NSGD \\
        \hline
         \multirow{5}{*}{Perfect} &\textbf{MQ2007} & 61.931 (5.535) & 59.936 (4.875) & 58.995 (4.926) & 59.765 (3.015)  & \textbf{68.639} (3.311) $\blacktriangle$\\
       & \textbf{MQ2008} & 81.327 (6.224) & 77.694 (6.137) & 76.192 (6.452) & 77.543 (4.827)  & \textbf{88.811} (6.022) $\blacktriangle$\\
       &\textbf{HP2003} & 110.012 (8.627) & 109.279 (8.565) & 92.422 (11.358) & 101.675 (4.943) & \textbf{113.890} (8.276) $\blacktriangle$\\
        &\textbf{NP2003} & 101.004 (8.702) & 98.774 (8.884) & 79.636 (13.338) & 104.677 (5.399)  & \textbf{115.145} (6.287) $\blacktriangle$\\
        &\textbf{TD2003} & 39.856 (7.770) & 38.054 (6.999) & 34.289 (7.703) & 38.380 (5.383)  & \textbf{42.402} (7.654) \\
        
        \hline
        \multirow{5}{*}{Navigational} &\textbf{MQ2007} & 57.989 (4.657) & 59.669 (4.911) & 57.301 (4.816) & 57.884 (3.266)  & \textbf{66.635} (2.832) $\blacktriangle$\\
        &\textbf{MQ2008} & 76.411 (5.983) & 75.603 (7.230) & 74.984 (5.959) & 75.001 (5.085)  & \textbf{84.091} (4.553) $\blacktriangle$\\
        &\textbf{HP2003} & 95.775 (14.394) & 95.925 (12.628) & 88.773 (11.518) & 82.244 (26.944) & \textbf{109.783} (5.634) $\blacktriangle$\\
        &\textbf{NP2003} & 84.699 (12.275) & 88.240 (13.039) & 74.521 (14.810) & 100.581 (8.962)  & \textbf{109.433} (5.649) $\blacktriangle$\\
        &\textbf{TD2003} & 33.954 (8.368) & 35.857 (8.729) & 31.468 (7.322) & 36.092 (5.616)  & \textbf{41.274} (7.318) $\blacktriangle$\\
        
        \hline
        \multirow{5}{*}{Informational}&\textbf{MQ2007} & 55.427 (5.639) & 57.094 (5.689) & 55.619 (5.066) & 55.338 (3.395)  & \textbf{67.312} (3.438) $\blacktriangle$ \\
        &\textbf{MQ2008} & 73.941 (6.101) & 74.825 (5.419) & 72.392 (6.259) & 72.757 (4.690)  & \textbf{84.053} (4.980) $\blacktriangle$ \\
        &\textbf{HP2003} & 59.376 (23.637) & 56.004 (22.101) & 66.295 (16.782) & 75.314 (11.281)  & \textbf{108.592} (5.503) $\blacktriangle$ \\
        &\textbf{NP2003} & 56.996 (20.547) & 54.615 (19.354) & 62.067 (17.667) & 74.497 (13.249) & \textbf{108.624} (5.831) $\blacktriangle$\\
        &\textbf{TD2003} & 23.021 (8.675) & 23.826 (7.964) & 24.948 (6.848) & 28.482 (5.299) & \textbf{39.386} (7.148) $\blacktriangle$\\
        \hline
    \end{tabular}
    \label{tab:online}
    \vspace{2mm}
    \caption{Offline score (NDCG@10) and standard deviation of each algorithm after 1000 queries under each of the three click models. Statistically significant improvements over MGD baseline are indicated by $\blacktriangle$ (p<0.05).}
    \vspace{-2mm}
    \begin{tabular}{cccccccccc}
        \hline
    Click Model   &Dataset& DBGD & CPS & DP-DBGD & MGD  & NSGD \\
        \hline
         \multirow{5}{*}{Perfect} &\textbf{MQ2007} & 0.369 (0.030) & 0.383 (0.026) & 0.361 (0.032) & 0.408 (0.018)  & \textbf{0.411} (0.019) \\
       & \textbf{MQ2008} & 0.465 (0.042) & 0.474 (0.042) & 0.461 (0.041) & 0.487 (0.037)  & \textbf{0.488} (0.043) \\
       &\textbf{HP2003} & 0.760 (0.067) & 0.764 (0.068) & 0.762 (0.062) & \textbf{0.771} (0.062) & 0.752 (0.752) \\
        &\textbf{NP2003} & 0.704 (0.052) & 0.702 (0.050) & 0.682 (0.062) & 0.712 (0.048)  & \textbf{0.714} (0.049)\\
        &\textbf{TD2003} & 0.267 (0.082) & 0.296 (0.094) & 0.286 (0.091) & \textbf{0.308} (0.096)  & 0.289 (0.092) \\

        \hline
        \multirow{5}{*}{Navigational}  &\textbf{MQ2007} & 0.359 (0.034) & 0.365 (0.037) & 0.339 (0.031) & 0.393 (0.024)  & \textbf{0.398} (0.022) \\
        &\textbf{MQ2008} & 0.459 (0.038) & 0.456 (0.037) & 0.445 (0.045) & 0.477 (0.036)  & \textbf{0.478} (0.037) \\
        &\textbf{HP2003} & 0.728 (0.063) & 0.734 (0.072) & \textbf{0.752} (0.061) & 0.707 (0.156)  & 0.744 (0.073) \\
        &\textbf{NP2003} & 0.709 (0.035) & 0.661 (0.066) & 0.675 (0.061) & 0.707 (0.052)  & \textbf{0.710} (0.039) \\
        &\textbf{TD2003} & 0.276 (0.095) & 0.285 (0.093) & 0.269 (0.087) & \textbf{0.303} (0.098)  & 0.274 (0.094) \\
       
        \hline
        \multirow{5}{*}{Informational}
        &\textbf{MQ2007} & 0.319 (0.047) & 0.325 (0.049) & 0.325 (0.037) & 0.355 (0.036)  & \textbf{0.383} (0.020) $\blacktriangle$\\
        &\textbf{MQ2008} & 0.425 (0.050) & 0.434 (0.047) & 0.422 (0.054) & 0.450 (0.041)  & \textbf{0.472} (0.036) \\
        &\textbf{HP2003} & 0.500 (0.196) & 0.463 (0.191) & 0.669 (0.103) & \textbf{0.736} (0.063)  & 0.713 (0.069) \\
        &\textbf{NP2003} & 0.526 (0.190) & 0.443 (0.179) & 0.657 (0.118) & 0.660 (0.059) & \textbf{0.707} (0.044) $\blacktriangle$\\
        &\textbf{TD2003} & 0.174 (0.099) & 0.178 (0.092) & 0.219 (0.094) & \textbf{0.271} (0.090) & 0.251 (0.085) \\
        \hline
    \end{tabular}
    \vspace{-3mm}
    \label{tab:offline}
\end{table*}

\subsection{Online and offline performance of NSGD}

We start with our first evaluation question: how does NSGD perform in comparison with baseline OL2R methods? We run all OL2R algorithms over all 5 datasets and all 3 click models. According to the standard hyper-parameter settings of DBGD \cite{yue2009interactively} and other baselines, we set $\delta$ to 1 and $\alpha$ to 0.1. For algorithms that can explore multiple candidates, including MGD  and NSGD, we set number of candidates explored in one iteration to 4, (i.e., $m = 4$ in NSGD). For NSGD, we set $k_g = 25$, $T_g = 15$, $k_h=10$, and $T_h = 50$. We will discuss the effect of these different hyper-parameters on NSGD in Section \ref{sec:hyperparameter}. All experiments are repeated 15 times for each fold, and we report the average performance.

Figures \ref{Fig:offline} and \ref{Fig:cumulative} report the offline and online performance of all OL2R methods on MQ2007 dataset under \emph{perfect}, \emph{navigational} and \emph{informational} click models. We also report the standard deviation of offline NDCG in every iteration of model update on this dataset in Figure \ref{Fig:std}. Due to the space limit, we cannot report the detailed performance over other datasets, but we summarize the final performance in Table \ref{tab:online} and \ref{tab:offline} respectively. From Figure \ref{Fig:cumulative}, we observe that CPS and NSGD, which both apply candidate preselection, perform better than other methods in terms of cumulative NDCG. This confirms that exploring carefully selected candidate directions generally improves the learning speed in the early iterations compared with the uniform sampling strategy used in other baselines. Our proposed NSGD further improves online learning efficiency over CPS by exploring inside the null space rather than the entire parameter space. From Table \ref{tab:online}, we can observe the consistent improvement of NSGD for most of the datasets and click models, which proves the accelerated learning speed by performing more efficient gradient exploration during its online learning process. 

In Figure \ref{Fig:offline} we first observe that NSGD improves offline NDCG significantly faster than other baselines, which generally require much more interactions with users to reach the same performance. This further explains our above analysis of the improved learning speed of NSGD shown in Figure \ref{Fig:cumulative}. For \emph{informational} users, MGD requires more than 800 iterations to reach performance comparable to NSGD at less than 200 iterations. From Table \ref{tab:offline} we observe that algorithms that explore multiple candidate directions in one iteration, including MGD and NSGD, consistently achieve better offline performance than other methods on all 5 datasets and 3 click models. Compared with MGD, NSGD further improves the final offline NDCG on MQ2007, MQ2008 and NP2003 datasets, especially for the \emph{informational} users. We have discussed in Section \ref{exp:setup} that MQ2007 and MQ2008 contain more queries with fewer assessments per query. This improvement suggests that NSGD can better identify the most effective exploration directions even under a \emph{noisy} environment. We have also tested MGD with 9 candidates explored in one iteration (i.e., $m=9$) which has achieves best performance according to \cite{schuth2016multileave}, and observed same consistent improvement of NSGD over MGD with 9 candidates in online performance. Due to space limit we did not report the performance of MGD with 9 candidates in Table\ref{tab:online} and \ref{tab:offline}.

Figure \ref{Fig:std} shows the standard deviation of offline NDCG at each iteration. We observe that both NSGD and MGD enjoy a much smaller standard deviation in the \emph{perfect} and \emph{navigational} users, suggesting that exploring multiple directions reduce the variance introduced by random exploration. Another reason for the reduced variance in NSGD is the hybrid sampling method mentioned in Section \ref{sec:nullspace}: the result confirms that first sampling from the basis vectors of null space and then sampling inside the null space provides a more effective exploration, which not only improves learning efficiency but also effectively reduces variance in an early stage. For \emph{informational} users, who have a lower stop probability and are likely to generate more clicks, they typically contribute nosier clicks and more ties in the comparison. In this case, NSGD reaches a much smaller standard deviation compared with MGD and all other baselines. The reason is that NSGD applies context-dependent candidate preselection to propose the most differentiable directions and use most difficult queries to discern tied candidates. Although CPS also uses historical interactions to preselect the rankers, it uniformly selects historical interactions, which are not necessarily informative. As a result, the ranking quality of CPS oscillates when the fidelity of user feedback is low.





\subsection{Zooming into NSGD}
\label{sec:hyperparameter}

To answer the second and third evaluation questions, we design detailed ablation studies to carefully study NSGD. All the experiments in this section were conducted on MQ2007 under the  \emph{informational} click model, as the dataset has the largest amount of queries and the click model makes the retrieval task the most challenging. 

\begin{figure*}[ht]
\centering
\vspace{-5mm}
\setlength\tabcolsep{3pt}
\begin{tabular}{ >{\centering\arraybackslash}m{5.7cm} >{\centering\arraybackslash}m{5.7cm} >{\centering\arraybackslash}m{5.7cm}}
\hspace*{0cm}
\includegraphics[width=5.8cm]{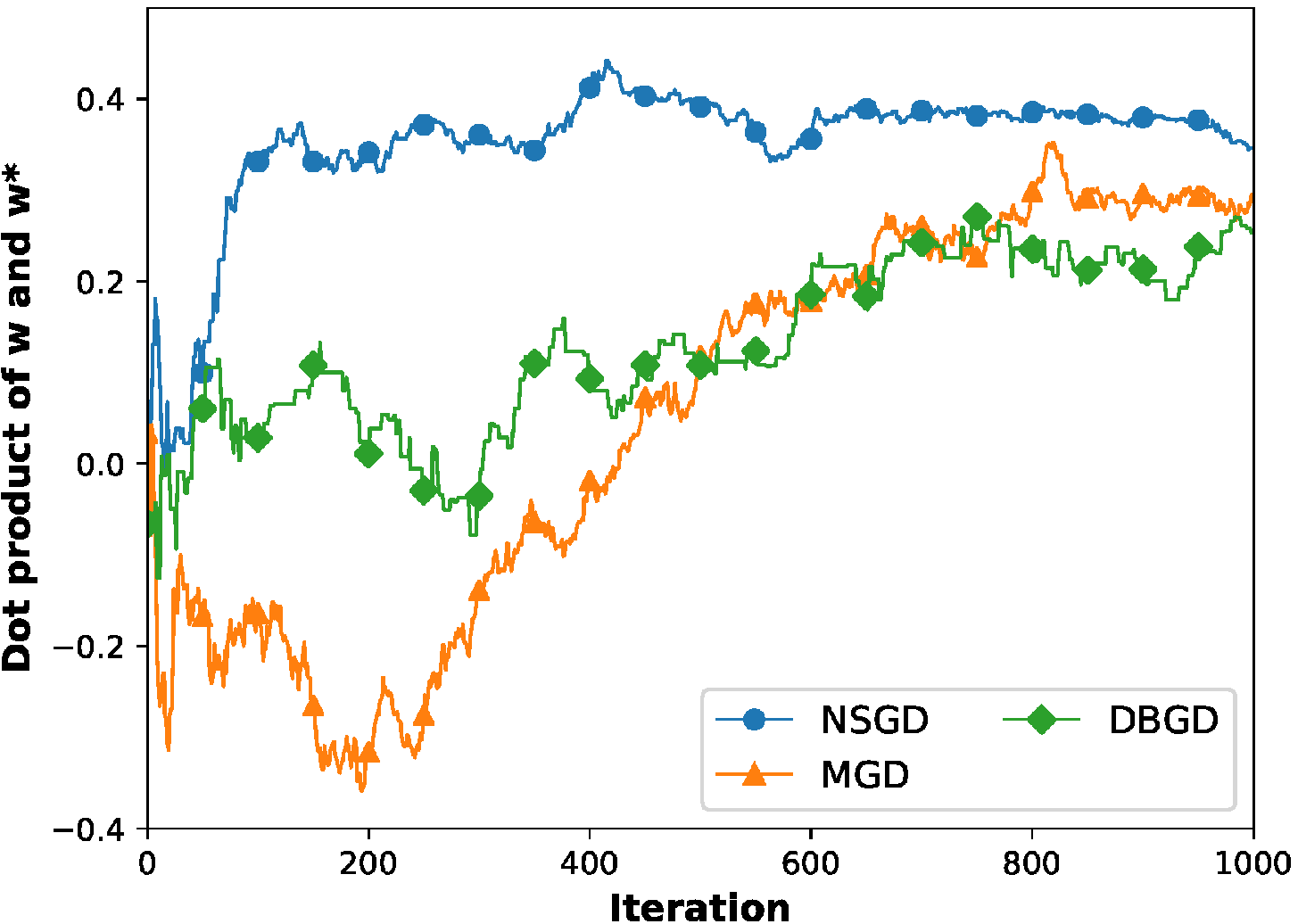} &
\includegraphics[width=5.8cm]{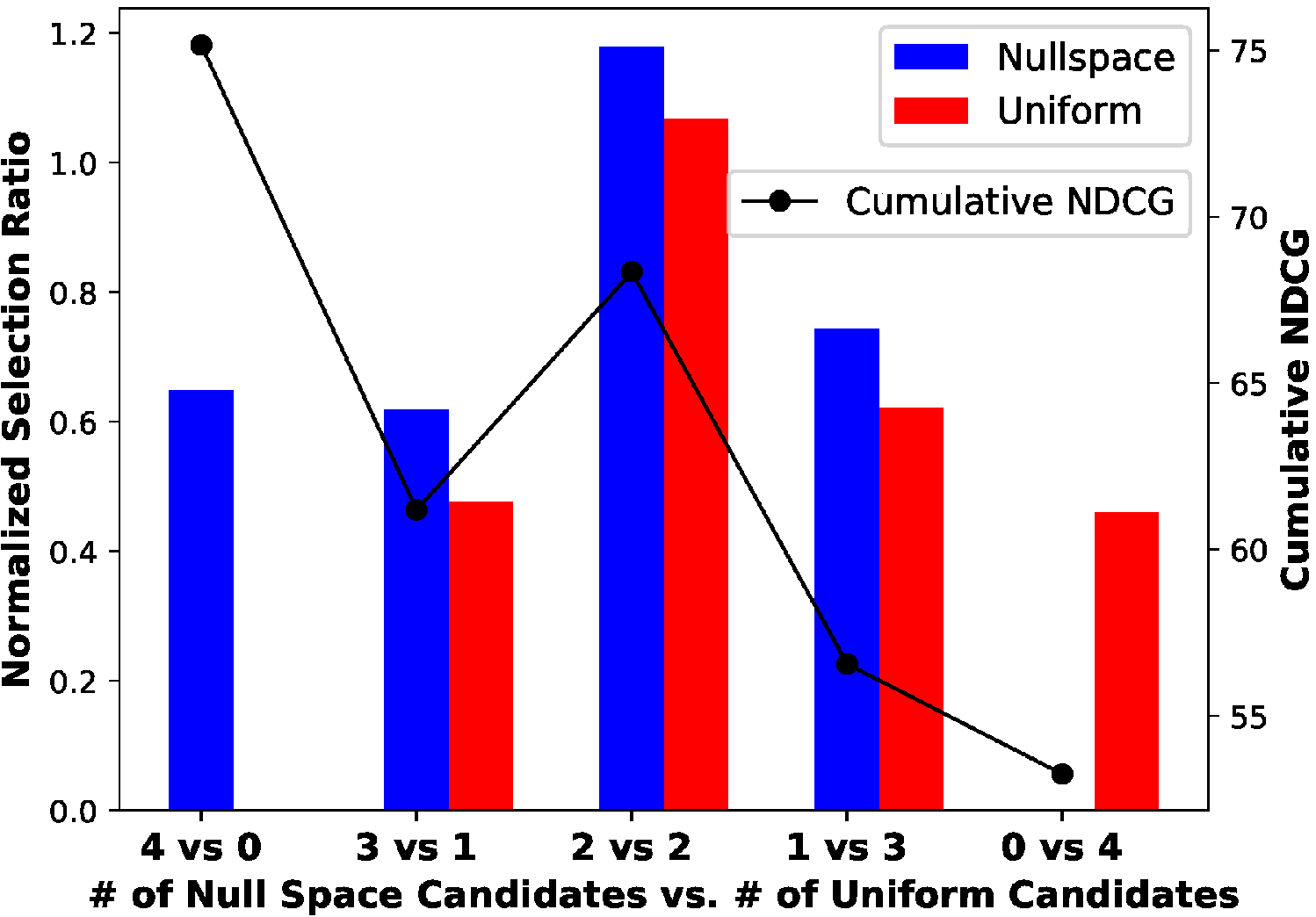} &
\includegraphics[width=5.8cm]{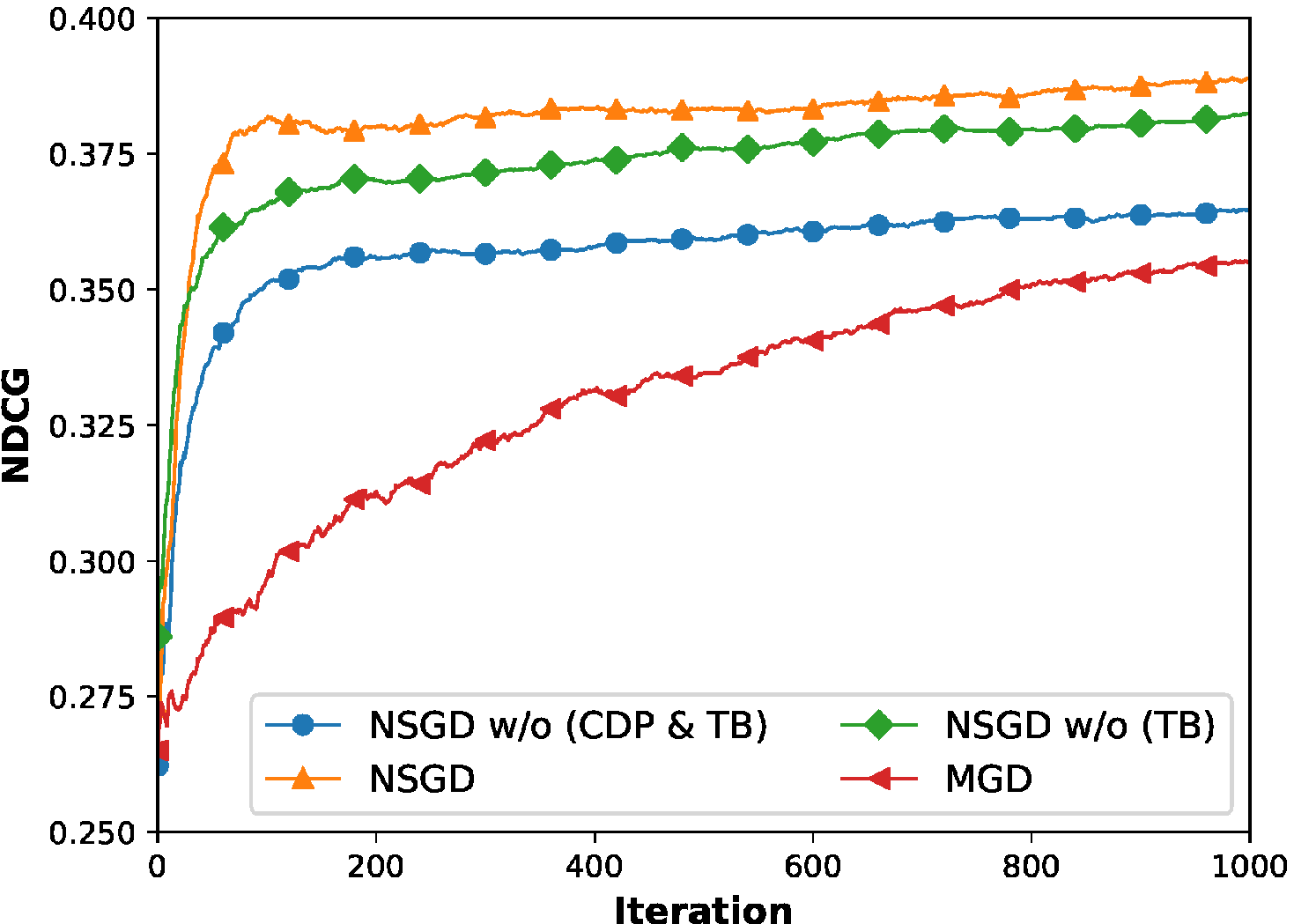} \\
(a) Cosine similarity between online learnt model and offline best model $w^*$ & (b)  Selection ratio comparing null space and uniform gradients & (c) Ablation analysis of NSGD  \\
\end{tabular}
\vspace{-3mm}
\caption{Detailed experimental analysis of NSGD on MQ2007 dataset.} \label{Fig:detail}
\vspace{-4mm}
\end{figure*}

In the first experiment, we trained an offline LambdaRank model \cite{burges2010ranknet} without any hidden layer using manual relevance labels. The model obtained the best offline NDCG performance in this dataset (around 0.437 in average). Its model parameter is denoted as $w^*$. We compare cosine similarity between the weight vector estimated by NSGD and $w^*$, to that between the weight vectors generated by MGD and DBGD and $w^*$ in each iteration. In Figure \ref{Fig:detail} (a) we can observe that NSGD moves towards $w^*$ much faster than both MGD and DBGD, which suggests the update directions explored by NSGD are more effective in recognizing the important ranking features. However, note that the final converged model in NSGD is not identical to $w^*$, and the final offline NDCG of all OL2R algorithms is worse than LambdaRank's. This is expected: LambdaRank is directly trained by manual labels. To improve an online trained model, one possible solution is to pre-train its weight vector with some offline data, and continue training it with online user feedback. This will take advantage of both training schemes. 

The second experiment serves to study the utility of gradients proposed by NSGD. We mix uniform exploratory directions from the entire parameter space with directions proposed from null space in the same algorithm. Specifically, we have in total 4 candidate rankers for multileaving, in which we vary the number of candidates created by null space gradients from 4 to 0, and we report the selection ratio, i.e., the frequency of selecting null space proposed rankers over the current ranker, versus the frequency of selecting uniformly proposed rankers over the current ranker. This ratio is also normalized by the number of proposed rankers in each type, to make the results comparable. We also report the online performance of each combination to understand the consequence of selecting different types of rankers. The result is shown in Figure \ref{Fig:detail} (b). We can clearly observe that comparing with the uniform exploratory rankers, rankers proposed by NSGD are always more likely to be selected as a better ranker in all combinations. We see that with more candidates proposed by NSGD, the online performance also increases. These results clearly show the superior quality of directions explored by NSGD, and explains the performance improvement observed comparing with other baselines that uniformly sample directions to explore.

To better understand the contribution of different components in NSGD, we disable them in turn and experiment on the resulting variants of NSGD. Specifically, we compare the following four models: 1) NSGD; 2) NSGD without tie breaking, denoted as NSGD w/o (TB); 3) NSGD without tie breaking and context-dependent preselection, denoted as NSGD w/o (CDP \& TB); and 4) MGD. The result is reported in Figure \ref{Fig:detail}(c). Comparing NSGD w/o (CDP \& TB) with MGD, where the difference is exploring in the null space or entire parameter space, confirms the utility of null space gradient exploration, which avoids repeatedly exploring recent and less promising directions. We want to mention that NSGD w/o (CDP \& TB) also significantly improves the learning speed and quickly reaches close to its high offline NDCG in less than 200 iterations, but it took MGD more than 800 iterations to achieve its highest performance. Comparing NSGD w/o (CDP \& TB) against NSGD w/o (TB), we observe that our context-dependent candidate preselection further improves the performance by selecting candidates that can be best differentiated by the current query in interleaved tests as compared with uniformly exploring inside the null space. Comparing NSGD w/o (TB) with NSGD, we observe that using difficult queries for tie breaking further improves the performance, rather than arbitrarily breaking the tie or taking the average of winners as suggested by \cite{schuth2016multileave}, which often introduces unexpected variance in online learning. 


\begin{figure*}[ht]
\centering
\vspace{-5mm}
\setlength\tabcolsep{0pt}
\begin{tabular}{ >{\centering\arraybackslash}m{4.3cm} >{\centering\arraybackslash}m{4.3cm} >{\centering\arraybackslash}m{4.5cm} >{\centering\arraybackslash}m{4.5cm}}

\hspace*{-3mm}
\includegraphics[width=4.4cm]{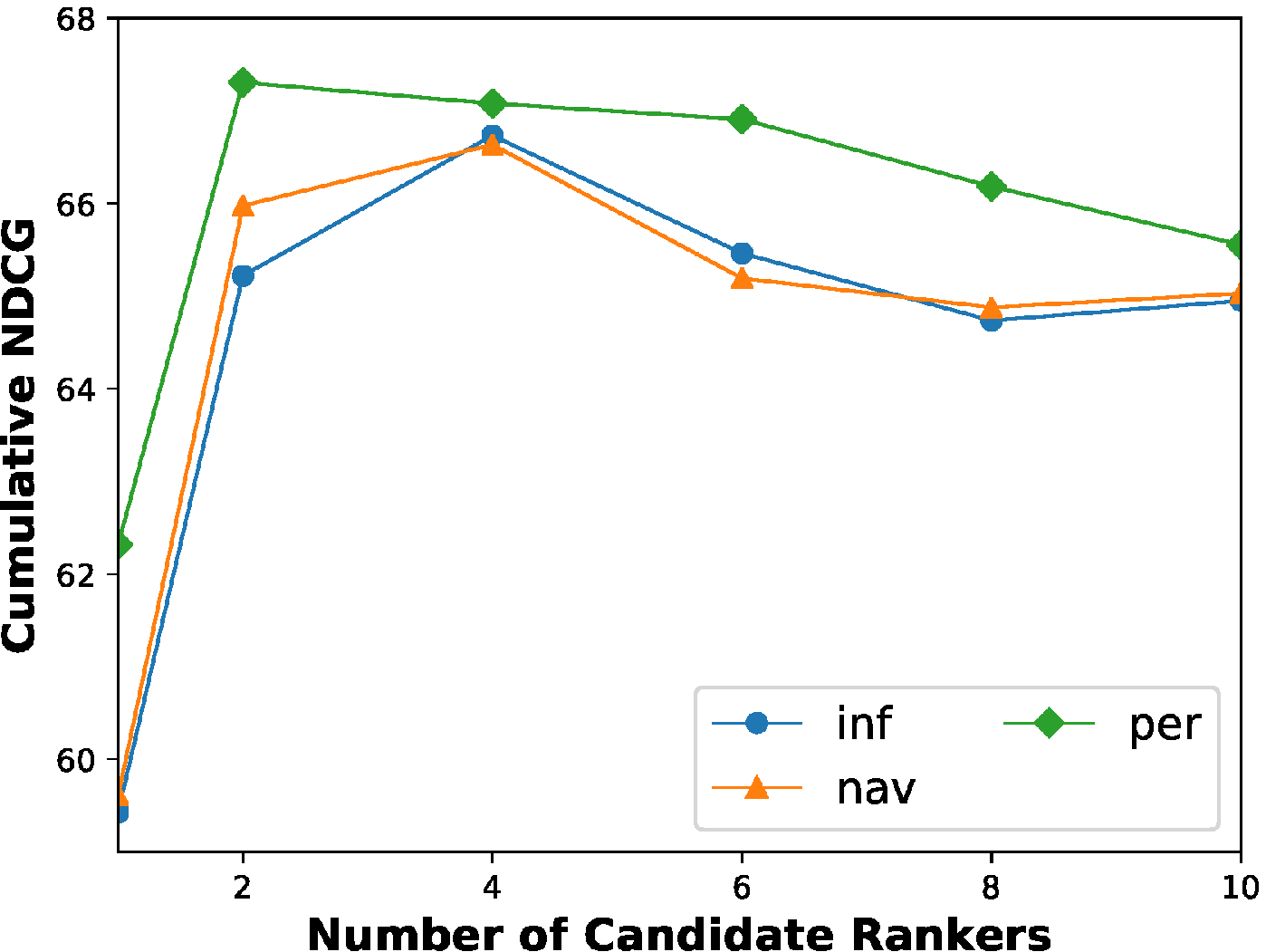} &
\includegraphics[width=4.4cm]{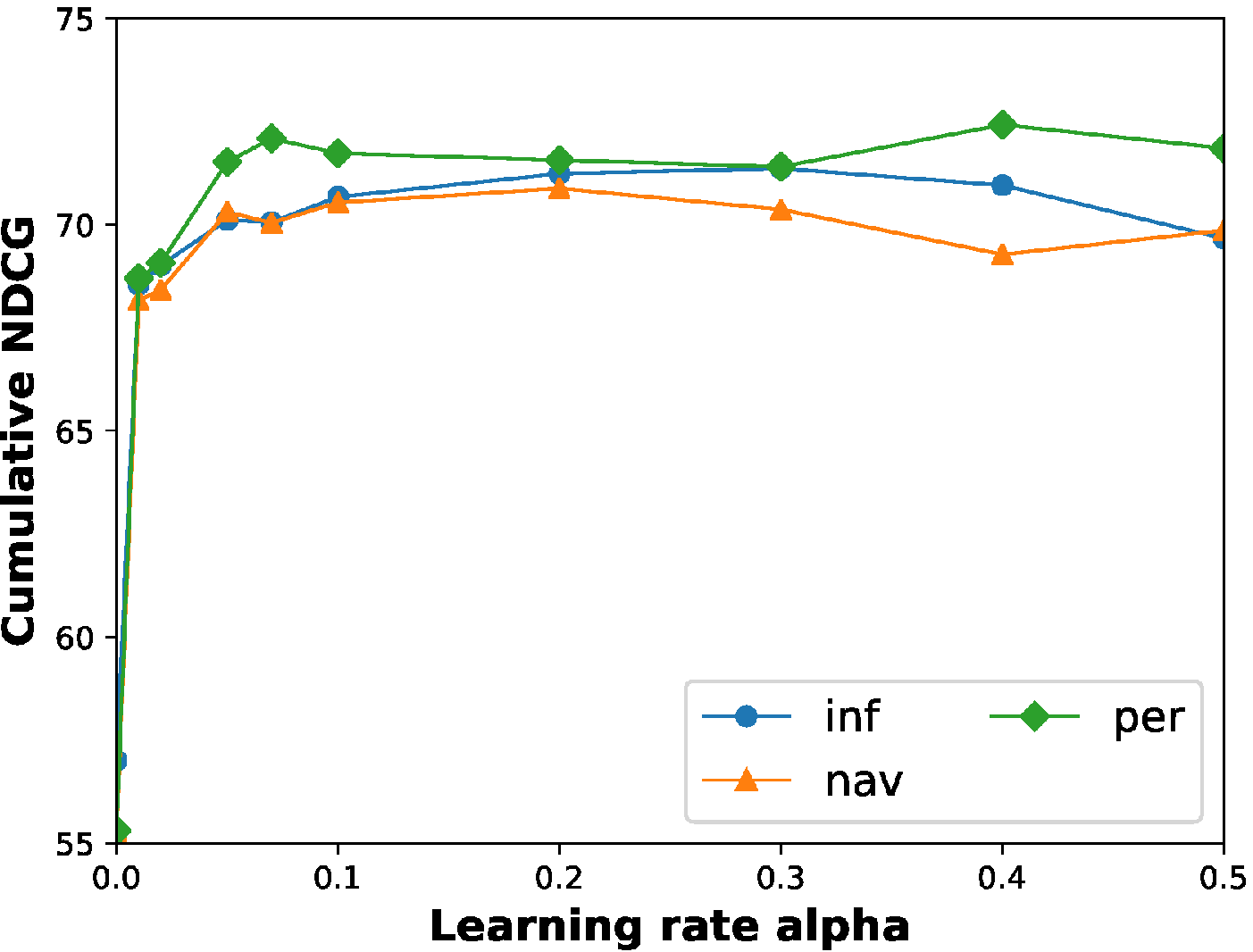} &
\includegraphics[width=4.4cm]{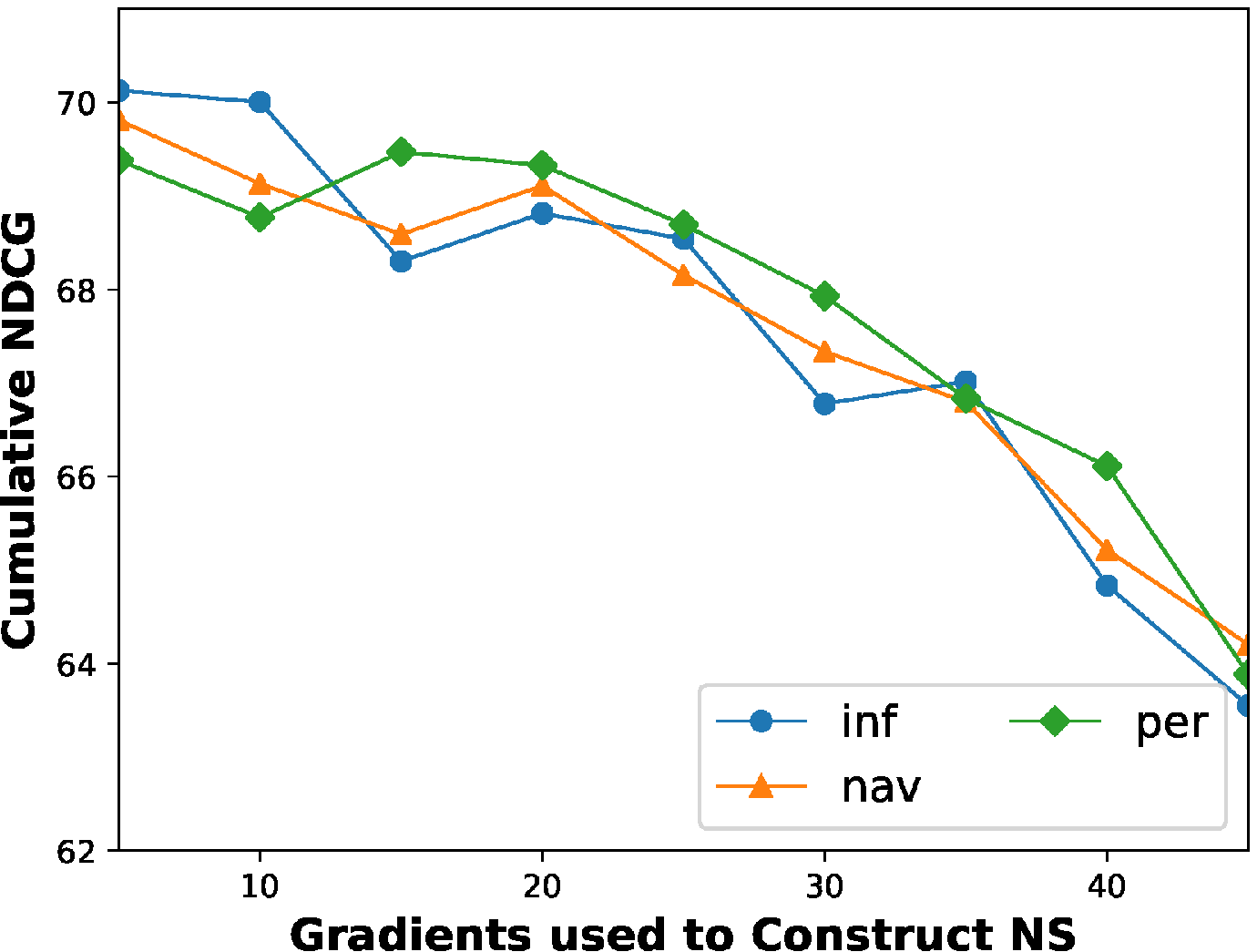} &
\includegraphics[width=4.4cm]{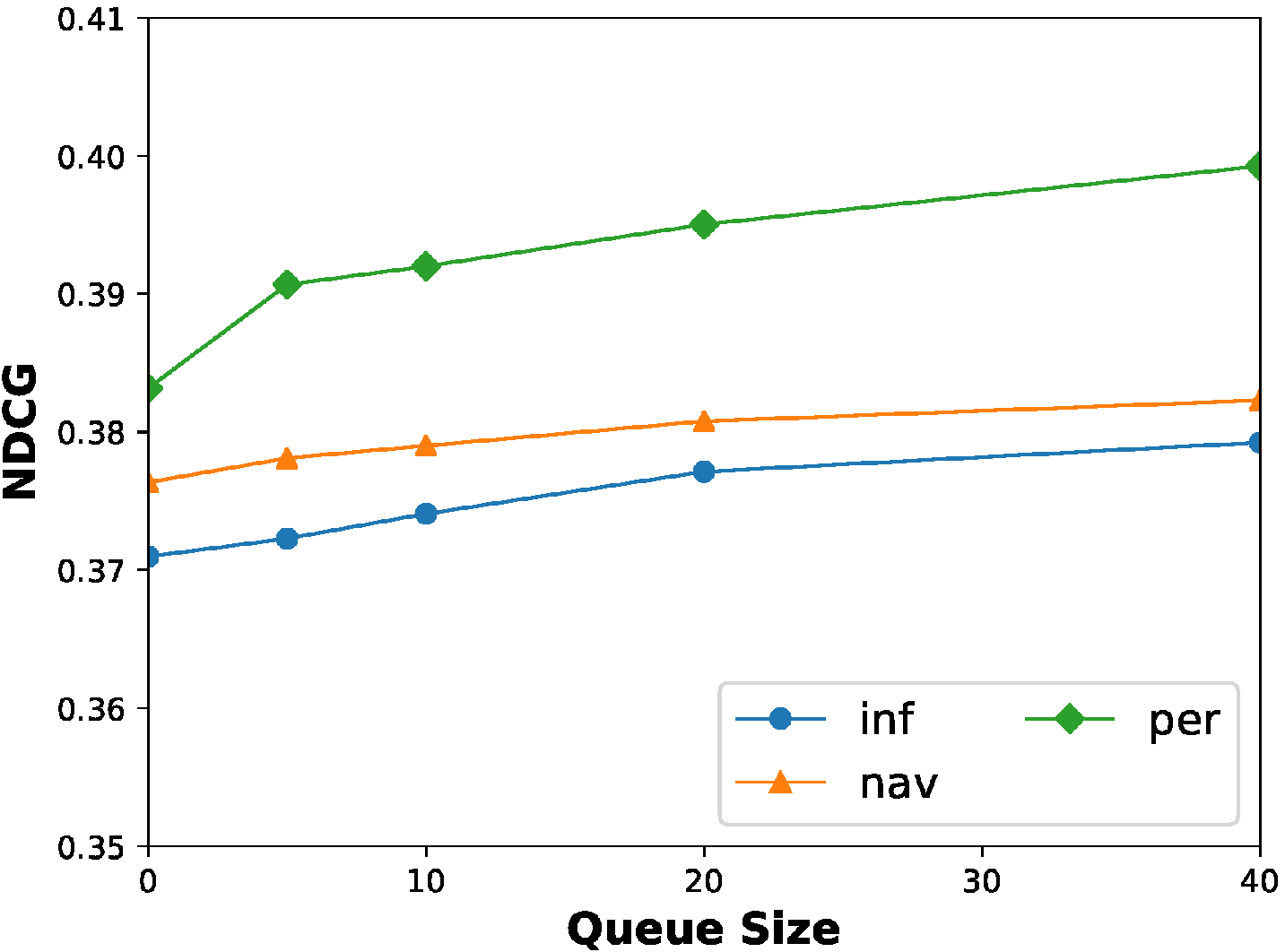} \\
(a) Number of candidates & (b)  Learning rate $\alpha$ & (c) Number of historical gradients to construct null space & (d) Number of historical queries for tie breaker\\
\end{tabular}
\vspace{-4mm}
\caption{Performance of NSGD under different hyperparameter settings on MQ2007 dataset.} \label{Fig:hyperparameter}
\vspace{-4mm}
\end{figure*}


To answer the fourth evaluation question, we study the effect of different hyper-parameter settings of NSGD and their corresponding online performance in the MQ2007 dataset with the three click models mentioned above.


\noindent\textbf{$\bullet$ Number of candidates.}  We vary the number of proposed candidate rankers $m$ from 1 to 10, from which the best ranker set is chosen through team-draft multileaving \cite{schuth2016multileave}. The result is reported in Figure \ref{Fig:hyperparameter} (a). Although each click model has different best-performing candidate size, with more candidate directions proposed the performance generally first increases and then slightly decreases. As more candidates are proposed, even though more directions can be explored, it is also easier to have multiple winners in the interleaved test, which introduces unnecessary complexity in recognizing the best ranker. For example, we can clearly observe a trend of decreasing performance across all click models when $m$ is larger than 5. Specifically, since the result list length is set to 10, each ranker will on average only receive 2 clicks when 4 new rankers are proposed. This makes it common to have tied winners. This serves as further motivation for having an effective tie-breaking function in NSGD. We do not present results for $m$ larger than 10 as each candidate ranker can only expect less than one click and the feedback from interleaved test is non informative.


\noindent\textbf{$\bullet$ Learning rate.} In all DBGD-type OL2R algorithms, the exploration step size is decided by $\delta$ and the learning rate for updating current ranker is decided by the choice of $\alpha$. Here we study the effect of different learning rate $\alpha$, by fixing $\delta$ to 1. Figure \ref{Fig:hyperparameter} (b) shows the result of varying $\alpha$ from 0 to 0.5. We notice that in most cases $\alpha$ around 0.1 gives the best performance. This suggests that even though we are exploring with a large step size $\delta$, we should use relatively small learning rate $\alpha$ to avoid over-exploration. 


\noindent\textbf{$\bullet$ Number of historical gradients to construct null space.} As mentioned in Section \ref{sec:nullspace}, when using $k_g$ historical gradients to construct the null space, NSGD only samples from a subspace whose rank is at most $d-k_g$. We vary the choice of $k_g$ from 5 to 40 (as in MQ2007 the feature dimension $d$ is 46). The result is showed in Figure \ref{Fig:hyperparameter} (c). We observe that when we increase $k_g$ to 20, the performance keeps relatively stable; but when $k_g$ goes beyond it, the performance decreases significantly. The key reason is the null space overly reduces the search space when $k_g$ is too large, such that it prevents NSGD from finding a good direction to explore, and forces it to converge to a suboptimal model quickly. 



\noindent\textbf{$\bullet$ Number of historical queries for tie breaking.} When the algorithm receives multiple winning rankers from an interleaved test, we use the most recent \emph{difficult} queries to identify the best ranker. In this experiment, we vary the number of historical queries $k_h$ for tie breaking from 0 (which means disable our tie breaking function) to 40. The result is showed in Figure \ref{Fig:hyperparameter} (d). Evidently, using more historical queries for tie-breaking leads to an increase the algorithm's performance. However, evaluating candidates over a large number of historical queries also increases the time and storage complexity. To balance computational efficiency and final performance, we set $k_h = 10$ for NSGD in all previous experiments.